\documentclass[preprint2]{aastex631} 
\usepackage{amsmath}
\usepackage{multirow}

\begin{document}

\title{The Star-forming Main Sequence and Bursty Star-formation Histories at $z>1.4$ in JADES and AURORA}

\author[0000-0003-1249-6392]{Leonardo Clarke}
\affiliation{University of California, Los Angeles, 475 Portola Plaza, Los Angeles, CA, 90095, USA}

\author[0000-0003-3509-4855]{Alice E. Shapley}
\affiliation{University of California, Los Angeles, 475 Portola Plaza, Los Angeles, CA, 90095, USA}

\author[0000-0001-9489-3791]{Natalie Lam}
\affiliation{University of California, Los Angeles, 475 Portola Plaza, Los Angeles, CA, 90095, USA}

\author[0000-0001-8426-1141]{Michael W. Topping}
\affiliation{Steward Observatory, University of Arizona, 933 N Cherry Avenue, Tucson, AZ 85721, USA}

\author[0000-0003-2680-005X]{Gabriel B. Brammer}
\affiliation{Cosmic Dawn Center (DAWN), Denmark}
\affiliation{Niels Bohr Institute, University of Copenhagen, Jagtvej 128, DK-2200 Copenhagen N, Denmark}

\author[0000-0003-4792-9119]{Ryan L. Sanders}
\altaffiliation{NHFP Hubble Fellow}
\affiliation{University of Kentucky, 506 Library Drive, Lexington, KY, 40506, USA}

\author[0000-0001-9687-4973]{Naveen A. Reddy}
\affiliation{Department of Physics \& Astronomy, University of California, Riverside, 900 University Avenue, Riverside, CA 92521, USA}

\author[0009-0002-6186-0293]{Shreya Karthikeyan}
\affiliation{University of California, Los Angeles, 475 Portola Plaza, Los Angeles, CA, 90095, USA}




\newcommand{\oiii}{[O\thinspace{\sc iii}]}
\newcommand{\neiii}{[Ne\thinspace{\sc iii}]}
\newcommand{\oii}{[O\thinspace{\sc ii}]}
\newcommand{\hii}{H\thinspace{\sc ii} }
\newcommand{\nii}{[N\thinspace{\sc ii}]}
\newcommand{\niii}{[N\thinspace{\sc iii}]}
\newcommand{\feii}{[Fe\thinspace{\sc ii}]}
\newcommand{\sii}{[S\thinspace{\sc ii}]}
\newcommand{\oi}{[O\thinspace{\sc i}]}

\newcommand{\secpoint}{\mbox{$''\mskip-7.6mu.\,$}}
\defcitealias{2023MNRAS.519.1526P}{P23}
\defcitealias{2014ApJS..214...15S}{S14}
\defcitealias{2024MNRAS.535.2998S}{S24}

\begin{abstract}

We analyze JWST spectroscopic and HST+JWST photometric observations of 659 star-forming galaxies at $1.4<z<9$ from DR3 of the JADES survey and the AURORA Cycle 1 program. We measure the star-forming main sequence (SFMS) for galaxies above $10^{8.5}\rm\ M_\odot$ where the sample is largely representative, estimating star-formation rates (SFRs) using the H$\alpha$ line flux and rest-frame far UV (1600\AA) continuum measurements, each independently corrected for dust attenuation. We find that the intrinsic, measurement-error-subtracted scatter in the SFMS ($\sigma_{\rm int}$) increases with decreasing stellar mass for the H$\alpha$-based SFMS, and we find no mass dependence of $\sigma_{\rm int}$ in the UV-based SFMS. Additionally, we find that $\sigma_{\rm int}$ decreases with increasing redshift, from $0.36^{+0.02}_{-0.02}$ dex to $0.22^{+0.08}_{-0.07}$ dex (H$\alpha$ SFMS), and from $0.28^{+0.02}_{-0.02}$ dex to $0.20^{+0.08}_{-0.07}$ dex (UV SFMS) between $z\sim2$ and $z\sim 6.5$. We also measure the redshift evolution of the specific SFR and find that, assuming $\rm sSFR\propto (1+z)^\gamma$, $\gamma=1.89^{+0.16}_{-0.15}$ for the H$\alpha$-based SFMS, and $\gamma=1.36^{+0.13}_{-0.13}$ for the UV-based SFMS. Analyzing the observed H$\alpha$/UV luminosity ratios and star-formation histories from the {\sc prospector} fitting code, we find that 41--60\% of the sample is inconsistent with having a constant star-formation history. Finally, we find tentative evidence for shorter SFR burst timescales with increasing redshift based on the distribution of $\rm L_{H\alpha}/\nu L_{\nu,1600}$ vs. $\Delta\rm \log(L_{H\alpha})$. Taken together, these results are consistent with recent theoretical predictions (e.g., the {\sc thesan-zoom} simulations) of bursty star formation in the early Universe and provide valuable constraints for theoretical models of galaxy evolution.

\end{abstract}



\section{Introduction} \label{sec:intro}

One of the goals at the forefront of modern galaxy evolution studies is to understand how the star-formation rates (SFRs) of galaxies in the early Universe evolve with cosmic time. Understanding the details of galaxy star-formation histories (SFHs) provides valuable insights into the physical processes that govern their growth. A key tool that has been used to study galaxy SFHs is the relationship between galaxy SFRs and stellar masses, commonly referred to as the star-forming main sequence \citep[SFMS;][also see \cite{2014ApJS..214...15S} and \cite{2023MNRAS.519.1526P} and references therein]{2004MNRAS.351.1151B,2004ApJ...617..746D,2007ApJ...660L..43N,2007A&A...468...33E,2007ApJS..173..267S,2010A&A...518L..25R,2012ApJ...754L..29W,2014ApJ...795..104W,2015ApJ...799..183S,2016ApJ...817..118T,2018A&A...615A.146P,2021MNRAS.505..540T,2021MNRAS.506.1237T,2024A&A...686A.128C,2024ApJ...972..156N,2024ApJ...977..133C,2025ApJ...979..193C,2025ApJ...981..161R,2025arXiv250804410S,2025arXiv250922871M,2025arXiv251000235W, perry2025prevalenceburstystarformation}. The SFMS is thought to arise primarily due to the fact that both stellar mass and baryon accretion rate scales with dark matter halo mass. Because the SFR increases with baryon accretion rate, there is a strong observable correlation between SFR and stellar mass. \citep[e.g.,][]{2013MNRAS.435..999D,2015ApJ...808...40W,2016MNRAS.455.2592R,2018ApJ...868...92T,2024ARNPS..74..173P}.

Observational constraints on the slope, normalization, and scatter of the SFMS over time serve as valuable benchmarks against which theoretical models and simulations can compare, thereby furthering our understanding of the growth of galaxies over time. The scatter around the SFMS in particular provides information regarding processes that affect galaxy SFHs on differing time scales. In this context, processes such as feedback from supernovae, ionizing radiation, and stellar winds represent short-timescale ($\lesssim$10 Myr) processes, while galaxy mergers and accretion on halo dynamical times represent long-timescale ($\gtrsim$100 Myr) processes \citep[e.g.,][]{2020MNRAS.498..430I,2023A&A...677L...4P,2025arXiv250300106M}. These processes are predicted to result in different SFMS scatter values as a function of stellar mass and redshift, depending as well upon which indicators are used to estimate the SFR. Indicators such as the Balmer-line luminosity (typically H$\alpha$) trace the ionizing radiation from massive O-stars, reflecting changes in the SFR on $\sim$5--10 Myr timescales. The non-ionizing far-UV (FUV) luminosity (1500\AA--1600\AA) is sensitive to longer-lived B-stars and is conventionally thought to trace changes in the SFR on $\sim$50--100 Myr timescales \citep{1998ARA&A..36..189K}.  However, simulation-based estimates of the true timescales probed by FUV radiation can vary between $\sim$10 Myr to $>$100 Myr, with increased timescales attributed to more bursty SFHs \citep{2021MNRAS.501.4812F}. Several observational studies have also used the ratio between the H$\alpha$ and the FUV luminosity to constrain the recent SFHs of galaxies. These studies compare the observed H$\alpha$/UV luminosity ratio with the predicted ratio based on a constant SFH and attribute large deviations from this predicted value to bursty star formation \citep[e.g.,][though see \citet{2023MNRAS.526.1512R} for a discussion on uncertainties in interpreting this ratio]{1999MNRAS.306..843G,2012ApJ...744...44W,2016ApJ...833...37G,2019ApJ...881...71E,2019ApJ...884..133F,2022MNRAS.511.4464A,2023ApJ...952..133M,2024MNRAS.52711372A,2024ApJ...977..133C,2025MNRAS.541.1348P}.

In general, high-resolution hydrodynamical simulations and several models of galaxy growth predict that lower-mass galaxies exhibit a larger degree of short-timescale stochasticity or ``burstiness" in their SFHs than high-mass galaxies due to the effects of stellar feedback in low-mass haloes as well as small numbers of giant molecular clouds \citep[e.g.,][]{2017MNRAS.466...88S,2020MNRAS.497..698T,2022MNRAS.511.3895F,2023MNRAS.525.2241H}. This burstiness is predicted to result in an increasing SFMS scatter with decreasing mass when the SFR is measured with short-timescale indicators such as H$\alpha$ \citep[e.g.,][]{2015MNRAS.451..839D,2017MNRAS.466...88S,2020MNRAS.498..430I}. Bursty SFHs are also expected to be ubiquitous among the high-redshift galaxy population, as gas accretion rates become highly variable, and galaxy dynamical timescales become shorter \citep[e.g.,][]{2017MNRAS.470.4698A,2018MNRAS.473.3717F,2018MNRAS.478.1694M,2020MNRAS.497..698T,2025arXiv250300106M}.

Observations with the James Webb Space Telescope (JWST) have continued to provide strong evidence that bursty SFHs play an important role in shaping the high-redshift galaxy population \citep[e.g.,][]{2024A&A...686A.128C,2024ApJ...977..133C,2025ApJ...979..193C,2025arXiv250616510M,2025arXiv250804410S,2025arXiv250713160C}. For instance, several studies have presented examples of galaxies that have undergone a rapid, dramatic decrease in their SFRs \citep[e.g.,][]{2023ApJ...949L..23S,2024Natur.629...53L,2025A&A...697A..90B,2025arXiv250622540C}, and a study by \citet{2024MNRAS.533.1111E} revealed signatures of rapidly rising and declining SFHs in the $6<z<9$ population through the distribution of emission-line equivalent widths. Bursty star formation has also been proposed as an explanation for the observed excess of UV-bright galaxies at $z>10$ discovered with JWST \citep[e.g.,][]{2023Natur.616..266L,2023ApJS..265....5H,2025arXiv250804791F}, as short-timescale fluctuations in the SFR can boost UV luminosities in low-mass galaxies \citep[e.g.,][]{2023MNRAS.521..497M,2023MNRAS.525.3254S,2023MNRAS.526.2665S,2024ApJ...975..192G}.

Observational studies measuring the scatter around the SFMS have typically found values in the range of 0.2-0.5 dex \citep[e.g.,][]{2014ApJS..214...15S,2015A&A...575A..74S, 2015ApJ...815...98S,2017ApJ...847...76S,2019MNRAS.483.3213P,2019MNRAS.490.5285P,2022ApJ...936..165L,2023MNRAS.519.1526P,2023ApJ...950..125M}, and studies with the JWST continue to measure values in this range \citep[e.g.,][]{2024ApJ...972..156N,2024ApJ...977..133C,2025ApJ...979..193C,2025arXiv250804410S,2025arXiv250922871M}. However, the literature has not yet reached a consensus on the mass and redshift dependence of the scatter around the SFMS \citep[][and references therein; \citealt{2024ApJ...977..133C,2025ApJ...979..193C,2025arXiv250804410S}]{2014ApJS..214...15S,2023MNRAS.519.1526P}. Recent works utilizing JWST/NIRCam observations have analyzed the SFMS scatter in large samples complete down to $10^{7.6-8.6}\rm\ M_\odot$ at $z>3$ \citep{2025ApJ...979..193C,2025arXiv250804410S,2025arXiv250922871M}. Currently, spectroscopic studies of the SFMS with JWST at $z\gtrsim2$ provide a complement to these photometric studies, but are limited in sample size and/or redshift range \citep[e.g.,][]{2024ApJ...977..133C,2024ApJ...972..156N,2024A&A...684A..75C,2025MNRAS.541.1348P,2025arXiv250708245T,perry2025prevalenceburstystarformation}. Thus, a large, representative sample of spectroscopically observed targets at $z\gtrsim2$ is necessary to supplement the existing photometry-based estimates of the SFMS and its scatter.

In this study, we quantify the burstiness of the galaxy population at $1.4<z<7$ using H$\alpha$-based and FUV-based estimates of the SFR, applying independent dust corrections to each quantity. We use data from the JWST Near-Infrared Camera \citep[NIRCam;][]{2023PASP..135b8001R} and Near-Infrared Spectrograph \citep[NIRSpec;][]{2022A&A...661A..80J} instruments taken as part of data release 3 
\citep[DR3;][]{2025ApJS..277....4D} of the JWST Advanced Deep Extragalactic Survey (JADES) program and Assembly of Ultradeep Rest-optical Observations Revealing Astrophysics (AURORA) GO program \citep{2025ApJ...980..242S}. Combining the JWST observations with Hubble Space Telescope (HST) data from the 3D-HST survey \citep{2012ApJS..200...13B,2014ApJS..214...24S}, we investigate how SFH burstiness varies as a function of galaxy stellar mass and redshift. In Section \ref{sec:obs}, we describe the observations, data reduction processes, and procedures for fitting SEDs and emission-line fluxes. Additionally, in Section \ref{sec:sample_analysis}, we compare the observed photometric properties of the combined JADES and AURORA sample to the larger representative photometric samples from 3D-HST \citep{2012ApJS..200...13B,2014ApJS..214...24S} and JADES \citep{2024MNRAS.535.2998S}. We present our measurements of the SFMS, the H$\alpha$/UV luminosity ratio distribution, and the sSFR distribution in Section \ref{sec:results}. In Section \ref{sec:discussion}, we discuss the implications of our results as they relate to the mass and redshift evolution of SFH burstiness. We summarize our conclusions in Section \ref{sec:conclusions}.

Throughout the paper, we assume the following cosmological parameters: $H_0 = 70\ \rm km\ s^{-1}\ Mpc^{-1}$, $\Omega_m = 0.3$, and $\Omega_\Lambda = 0.7$. For the initial mass function (IMF), we assume the \citet{2003PASP..115..763C} form. For chemical abundances and Solar abundance patterns, we use the \citep{2009ARA&A..47..481A} values, with $\rm 12+\log{(O/H)}_\odot = 8.69$, and $\rm Z_\odot = 0.014$.

\section{Observations and Data Reduction} \label{sec:obs}

\subsection{JADES}

\subsubsection{NIRCam and HST imaging}
The majority of the dataset in this study consists of photometric and spectroscopic measurements from DR3 of the public JADES GTO and GO programs (PIDs: 1180, 1181, 1210, 1286, 3215, \citealt{2023arXiv230602465E,2025ApJS..277....4D}). Data from the High-Level Science Project JADES were obtained from the Mikulski Archive for Space Telescopes (MAST) at the Space Telescope Science Institute (STSCI) \citep{https://doi.org/10.17909/8tdj-8n28}. \mbox{
}\hspace{0pt}
The associated photometric catalogs that we used in this study, which are described by \citet{2023ApJS..269...16R}, \citet{2023arXiv230602465E,2023arXiv231012340E}, and \citet{2024ApJ...970...31R}, are available on the JADES MAST page\footnote{https://archive.stsci.edu/hlsp/jades}. The JADES dataset includes galaxies in both the GOODS-N and GOODS-S extragalactic legacy fields.

The JADES photometric catalog utilized JWST/NIRCam imaging taken as part of the program's primary and parallel observations in several filters, including F070W, F090W, F115W, F150W, F200W, F277W, F335M, F356W, F410M, and F444W. Supplementing the JADES imaging are NIRCam observations taken with the F182M, F210M, and F444W filters from FRESCO \mbox{
\citep{2023MNRAS.525.2864O}}\hspace{0pt}
, and F182M, F210M, F430M, F460M, and F480M data from JEMS \mbox{
\citep{https://doi.org/10.17909/fsc4-dt61,2023ApJS..268...64W}}\hspace{0pt}
. \citep{2023ApJS..268...64W}. Data from the High-Level Science Projects FRESCO and JEMS were obtained from the MAST archive at STScI \citep{https://doi.org/10.17909/gdyc-7g80,https://doi.org/10.17909/fsc4-dt61}. Finally, the photometric catalog utilizes HST/ACS and HST/WFC3 mosaics in the GOODS-N and GOODS-S fields \citep{2013ApJS..209....6I,2019ApJS..244...16W}, adding photometry in the HST/ACS F435W, F606W, F775W, F814W, and F850LP filters as well as HST/WFC3 F105W, F125W, F140W, and F160W.  For the purposes of this study, we use the \texttt{KRON\_CONV} extension of the MAST photometric catalog, which corresponds to photometric measurements that have been matched to the NIRCam F444W point-spread function (PSF) and measured using \cite{1980ApJS...43..305K} elliptical apertures.

\subsubsection{NIRSpec}

The JADES/NIRSpec Micro-shutter Assembly \citep[MSA;][]{2022A&A...661A..81F} observations are presented in \citet{2024A&A...690A.288B} and \citet{2025ApJS..277....4D}. We briefly describe the spectroscopic data and their reduction here. The NIRSpec/MSA observations consist of low-resolution ($R\sim 100$) prism as well as both medium- ($R\sim 1000$) and high-resolution ($R\sim2700$) grating spectra. Of these, we utilized the prism and the medium-resolution grating data. The medium-resolution data were taken with three grating/filter combinations: G140M/F070LP, G235M/F170LP, and G395M/F290LP. The observations were taken in three categories, or ``tiers" of exposure depth: medium (3,588 objects), deep (253 objects), and ultradeep (228 objects). For the prism observations, the average exposure times for the medium, deep, and ultradeep tiers were 1.6 h, 16.5 h, and 32.4 h, respectively. For the medium-resolution grating observations, the average exposure times for the medium, deep, and ultradeep tiers were 1.3 h, 4.1 h, and 15.6 h (7.7 h in G140M, 23.0 h in G395M, and no observations in G235M for the ultradeep tier), respectively. The detailed breakdown of exposure times for each tier can be found in \citet{2025ApJS..277....4D}. Targets were observed through three-microshutter slitlets on the MSA, utilizing a three-point nod pattern between microshutters, which offsets exposures by $0 \secpoint 529$ along the long axes of the slits. 

We downloaded the version 3 reduced and extracted 1D spectra from the Dawn JWST Archive (DJA)\footnote{https://dawn-cph.github.io/dja/}. The data reduction process is described by \citet{2025A&A...697A.189D} and \citet{2025A&A...693A..60H}. In brief, the reduction follows the standard \texttt{jwst}\footnote{https://jwst-pipeline.readthedocs.io/en/latest/index.html} STScI reduction pipeline, processing the raw, uncalibrated frames through the Stage 2 \texttt{calwebb\_spec2} routine. Beyond this stage, the data were calibrated using \texttt{msaexp} \citep{https://doi.org/10.5281/zenodo.7299500} to combine the exposures from the different nod positions, producing rectified, background-subtracted, two-dimensional (2D) spectra. Each object's spatial profile was then automatically fit with a Gaussian profile in order to define the traces, and one-dimensional (1D) spectra were then optimally extracted \citep{1986PASP...98..609H} from these 2D frames.

For those objects more spatially extended than the $\sim$$0\secpoint5$ nods, we modified the standard data reduction to remove the middle nod position, thereby minimizing self-subtraction when combining the dithered exposures. We elaborate on this process in Appendix \ref{sec:ext_objs}. In total, we identified and re-reduced 115 extended objects, 46 of which are included in the SFMS sample (see Section \ref{sec:selection} for sample description).

For the G140M/F070LP grating observations, because the spectra provided by the DJA truncate at 1.25$\rm \mu m$ to avoid contamination from higher-order dispersion in the range $\rm \sim1.25-1.8\ \mu m$, we instead utilize the G140M/F070LP grating observations provided by the JADES team, available on their website\footnote{https://jades-survey.github.io/scientists/data.html}, since their reductions extend up to 1.8$\rm \mu m$. We describe our approach for dealing with contamination from higher-order dispersion in Section \ref{sec:linefitting}.

\subsubsection{Spectroscopic Flux Calibration} \label{sec:fluxcal}

Because the design of the NIRSpec/MSA for JADES allowed spectral traces to overlap \citep[see][]{2025ApJS..277....4D}, many of the grating spectra, whose traces subtend more of the detector than do the prism spectra, dispersed on top of each other, leading to contaminated spectra. If unaccounted for, this contamination would lead to inaccuracies in the flux calibration when comparing the grating spectra to the photometry. Since the traces of the prism spectra are shorter, they did not suffer from this contamination issue to the same extent. For this reason, we chose to only match the prism spectra to the photometry and used the grating spectra for the purpose of resolving closely spaced emission lines such as H$\alpha$ and \nii $\lambda \lambda 6550,6585$ (see discussion in Section \ref{sec:linefitting}).

To ensure that the flux calibration of the prism spectra was consistent with the photometry, we first created synthetic photometric data points by passing each spectrum through the photometric filter curves that overlapped it in wavelength. We only included filters whose observed {\it and} synthetic photometry were both measured with $>$$3\sigma$ significance. Then, using the ratio of the synthetic over the observed photometric flux density, we calculated a multiplicative scaling factor for each filter. We fit this set of multiplicative scaling factors with a wavelength-dependent polynomial of the form:

\begin{equation}\label{eq:fluxcal}
    s(\lambda) = \sum_{i=0}^4 c_i \lambda^i
\end{equation}

where $c_i$ is the $i$-th order coefficient, and $\lambda$ is the wavelength in Angstroms.

 To create the flux calibration polynomial for each prism spectrum, we first fit the set of scaling factors with a 0th- ($c_{i>0} =0$) and a 1st-order ($c_{i>1} =0$) polynomial and performed a statistical $F$-test \citep{hahs2020introduction}, comparing the variances of the residuals to determine which order polynomial was preferred. If the 0th order was preferred, the 0th-order fit became our fiduciual flux calibration function. If the 1st order was preferred, we then performed an $F$-test comparing the 1st- ($c_{i>1} =0$) and 2nd-order ($c_{i>2} =0$) fit residuals and correspondingly adopted the better of the two polynomial fits. In the case of the 1st-order fit, if the $F$-test indicated that a 1st-order polynomial was preferred compared to a 2nd order, we also checked higher orders beyond the 2nd order to ensure the most robust fit. We performed this polynomial analysis with a 4th-order polynomial ($i=4$) being the highest order considered. The final flux-calibrated spectra were then calculated as $f_{final} = f_{initial} \times s(\lambda)$. Of the 625 JADES galaxies considered in our SFMS analysis, our flux calibrations consisted of 334 0th-order, 189 1st-order, 78 2nd-order, 22 3rd-order, and 2 4th-order scaling polynomials.

In order to mitigate the effects of unrealistic end behavior for the polynomial fits in wavelength regions that were unconstrained by the photometry, we truncated the polynomial fits at 1000 \AA\ bluewards and redwards of the bluest and reddest detected photometric data points, respectively. We then extrapolated the scaling factor at the bluest end of this truncated polynomial as a constant value to the end of the wavelength coverage of the prism, and we did the same for the reddest end of the truncated polynomial. We present a summary plot of the resulting flux calibration in Appendix \ref{sec:flux_cal_app}. 

\subsubsection{SED fitting}

We fit the spectral energy distributions (SEDs) of our objects using the {\sc prospector} SED-fitting code \citep{2021ApJS..254...22J}. We fit galaxies assuming a non-parametric star-formation history (SFH), adopting a Student-t distribution as a continuity prior, breaking up the SFH into eight bins in cosmic time \citep[see, e.g.,][]{2022ApJ...927..170T}. We fixed the two most recent time bins to encompass 3 Myr and 10 Myr of cosmic time, respectively, and evenly spaced the six remaining time bins logarithmically in time back to $z=20$. We ran {\sc prospector} utilizing Flexible Stellar Population Synthesis \citep[FSPS;][]{2009ApJ...699..486C, 2010ApJ...712..833C} as the stellar evolution model, which uses the MILES stellar spectral library \citep{2006MNRAS.371..703S} and MIST isochrones \citep{2011ApJS..192....3P,2013ApJS..208....4P,2015ApJS..220...15P,2016ApJS..222....8D,2016ApJ...823..102C}. For each object, we used the multi-wavelength photometry during fitting. Emission-line fluxes were tied to the best-fit {\sc cloudy} \citep{2013RMxAA..49..137F} model, drawn from a pre-computed grid \citep{2017ApJ...840...44B}.

Following earlier work \citep[e.g.][]{2024ApJ...977..133C,2025MNRAS.541.1707T}, we fit the SED of each object with two sets of assumptions for the metallicity and the dust attenuation law. For one set, which we refer to as ``SMC+0.28Z$_\odot$,'' we fit the SED assuming an SMC \citep{2003ApJ...594..279G} dust attenuation law, and we fixed the metallicity of the stars and gas during the fit to be 28\% of the Solar value. In the other assumption scheme, which we refer to as ``Calz+1.4Z$_\odot$,'' we fit the SED assuming a \citet{2000ApJ...533..682C} dust attenuation law, and we fixed the metallicity of the stars and the gas to be 140\% of the Solar value. For each object, to choose the best SED fit between the two, we compared the output maximum probability from {\sc prospector}, and we adopted the fit with the greater probability of the two to be our fiducial fit. When comparing the stellar masses determined with either set of assumptions, the median and standard deviation of the difference between the two estimates ($\rm \log(M_*/M_\odot)_{Calz+1.4Z_\odot} - \log(M_*/M_\odot)_{SMC+0.28Z_\odot}$) is $0.05\pm0.26$, showing that choosing between the two assumptions does not systematically bias the stellar masses.

\subsubsection{Emission line fitting}\label{sec:linefitting}

Due to the presence of overlapping grating spectra in the MSA grating observations as well as the presence of higher-order dispersion in the range of 1.25--1.8 $\rm \mu m$ in the G140M/F070LP spectra, we chose to fit emission-line fluxes simultaneously in the prism and the medium-resolution grating spectra. With this approach, we used the flux-calibrated prism spectra to fit for the absolute fluxes of emission lines while simultaneously fitting the same lines in the grating spectra to resolve closely spaced emission lines (e.g., H$\alpha$ and \nii). As a result, we allowed the integrated fluxes of emission lines to be different in the prism vs. the grating for the same object, while still preserving the flux ratios of all emission lines (for example, we ensure that $f_{\rm [N II]} / f_{\rm H\alpha}$ remains the same in both the prism and grating during fitting. In the particular case of \nii/H$\alpha$, we weight the likelihood function such that the ratio is constrained by the grating only). In summary, the emission-line fluxes reported and analyzed in this study are based on the fits to the lines in the flux-calibrated prism spectra, with constraints on the relative line flux ratios provided by simultaneous fits to the lines in the gratings.

In order to remove spurious emission lines arising from contamination due to light from adjacent slits, we visually inspected the prism and grating spectra of each object side-by-side. Upon visual inspection, we manually masked out contaminating features in the grating spectra that were not present in the prism. 

To secure the redshift during emission-line fitting, we first performed a preliminary measurement of only the brightest emission lines using \texttt{lmfit} \citep{https://doi.org/10.5281/zenodo.15014437}, assuming Gaussian profiles and flat continua. We then used the redshift from this initial fit to proceed to the main, custom emission-line fitting routine.

To model the continuum during the main fitting procedure, we used the sum of the stellar and nebular continuum emission spectrum from the best-fit {\sc prospector} SED model. For the prism spectra, we smoothed the continuum model to match the resolution of the prism at each wavelength. Each emission line was modeled with a Gaussian line profile on top of the continuum model. 

The width of each emission line was calculated to be the width of the instrumental resolution at the wavelength of the given line convolved with the line velocity dispersion, which was included as a parameter during fitting. We modified the NIRSpec instrumental resolution curves from STScI\footnote{https://jwst-docs.stsci.edu/jwst-near-infrared-spectrograph/nirspec-instrumentation/nirspec-dispersers-and-filters\#gsc.tab=0} by multiplying them by a factor of 1.5. The STScI resolution curves assume a uniformly illuminated slit \citep[see discussion in][]{2024A&A...684A..87D}, however, many JADES targets are compact or point sources, resulting in a higher instrumental resolution than predicted. These compact objects required a smaller lower limit on the line spread function width than what would be allowed given the provided resolution curves, and we found that a factor of 1.5 gave a better match to the data.

Using Gaussian emission profiles superimposed on the continuum as the model in our likelihood function, we constructed a prior function, limiting the line fluxes to be positive values. We sampled the resulting probability function using the Python Markov-Chain Monte Carlo (MCMC) package \texttt{emcee} \citep{2013PASP..125..306F}. We checked for convergence every 5,000 steps for a maximum of 100,000 steps, stopping the sampling if the number of steps exceeded 30 times the median autocorrelation time for the parameters and changed by less than 5\% from the last evaluation. For objects in the JADES medium-depth tier, we estimate 3$\sigma$ line flux limits of $5.4\times10^{-19}\rm\ erg\ s^{-1}\ cm^{-2}$, $3.2\times10^{-19}\rm\ erg\ s^{-1}\ cm^{-2}$, and $2.2\times 10^{-19}\rm\ erg\ s^{-1}\ cm^{-2}$ in the wavelength ranges of the G140M, G235M, and G395M gratings, respectively.

\subsection{AURORA}

In addition to the JADES data, we included observations obtained as part of the AURORA GO program \citep[PID: 1914,][]{2025ApJ...980..242S}. Data from the High-Level Science Project AURORA were obtained from the MAST archive at STSCI
\citep{https://doi.org/10.17909/evwk-m381}. These data consist of deep NIRSpec MSA observations in the GOODS-N and COSMOS fields taken with the medium-resolution grating configuration, continuously covering $1-5\rm\ \mu m$ across the G140M/F100LP, G235M/F170LP, and G395M/F290LP grating/filter combinations. Below, we describe the reduction of the NIRSpec observations as well as the photometry and SED fitting procedures.

\subsubsection{Photometry and SED fitting}

The AURORA targets are located in the COSMOS and GOODS-N extragalactic legacy fields, for which a large number of HST and JWST filters is available from JADES, FRESCO \citep{2023MNRAS.525.2864O}, PRIMER (Dunlop et al., in prep; \citet{2024MNRAS.533.3222D}), and 3D-HST \citep{2012ApJS..200...13B,2014ApJS..214...24S}. The details of the photometry used in the processing and analysis of AURORA data are described in \citet{2025ApJ...980..242S}.

We utilize the SED fits detailed in \citet{2025MNRAS.541.1707T}, which were generated using the {\sc prospector} code \citep{2021ApJS..254...22J} assuming an eight-bin non-parametric SFH, with the most recent SFH bin fixed to 10 Myr in duration and the remaining bins evenly logarithmically spaced to the age of the Universe. Prior to fitting, the photometric data were corrected for contributions from strong emission lines as well as for nebular continuum emission calculated based on the strengths of the detected Balmer lines. The SEDs were fit with the same metallicity and dust-law pattern that we adopted in this work, choosing the lowest $\chi^2$ fit as the fiducial SED model for each galaxy. We also compared the stellar mass estimates from \citet{2025MNRAS.541.1707T} estimated by simultaneously modeling the nebular and stellar emission in the photometry without prior emission-line flux corrections. The median and standard deviation of the difference between these two stellar mass estimates for the sample is $0.05\pm 0.34$, showing that differences in accounting for emission-line contributions does not systematically bias the stellar mass estimates.

\subsubsection{NIRSpec data reduction}

The NIRSpec data reduction for the AURORA spectra is described in \citet{2025ApJ...980..242S}. From the reduced 2D spectra, the 1D spectra were slit-loss-corrected as outlined in \citet{2023ApJ...948...83R}, and the spectra were flux-calibrated in two stages: the first stage correcting band-to-band relative flux calibration as in \citet{2024ApJ...962...24S}, and the second stage scaling the spectra to match the available photometry. The emission lines were modeled as Gaussian profiles, and the fitting proceeded in two stages. The first stage involved using the non-emission-line-corrected best fit SED from {\sc fast} \citep{Kriek_2009} to model the continuum during line fitting. After the initial line fit, these line fluxes were subtracted from the photometry along with the nebular continuum emission inferred from the Balmer lines. The final line fluxes were fit using the nebular-emission-corrected photometry, and a total of 95 galaxies yielded significant emission-line detections. Because the galaxies at $z\gtrsim3$ are extreme objects with large specific SFRs (sSFRs), we only include the galaxies in the redshift range $1.4<z<2.7$, which have properties more representative of the galaxy population at these redshifts, adding a total of 40 objects to the analysis.

\subsection{Combined Spectroscopic Sample}

\begin{figure}
    \centering
    \includegraphics[width=8.6cm]{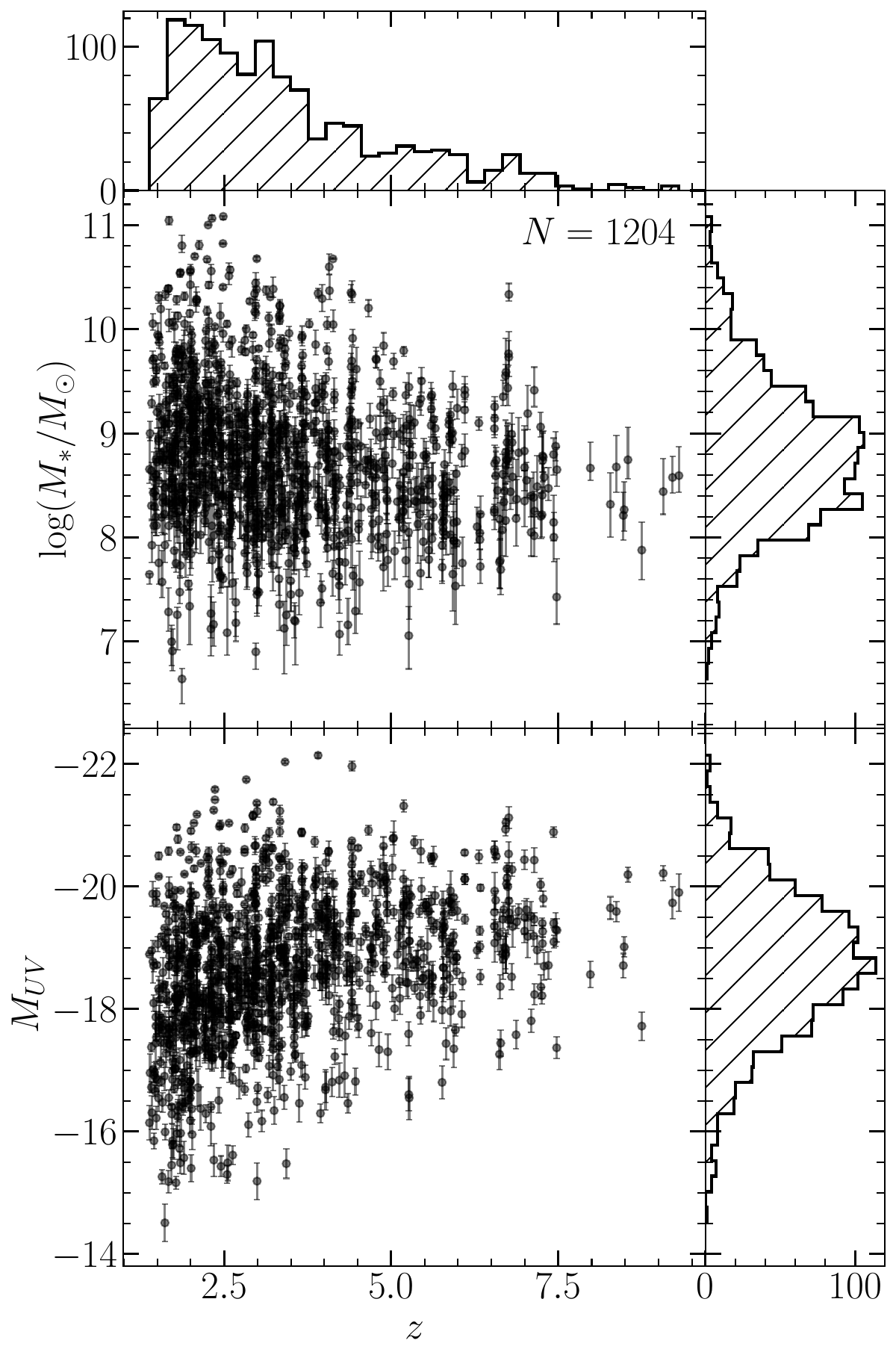}
    \caption{Stellar mass and UV magnitude distributions vs. redshift for the JADES and AURORA spectrscopic sample.}
    \label{fig:zdist}
\end{figure}

The spectroscopic sample that we analyze in this study is drawn from the larger NIRSpec sample from the JADES and AURORA surveys. We will first describe the spectroscopic sample from JADES, which consists of 4,086 objects, as detailed in \citet{2025ApJS..277....4D}. Of the 4,086 objects, we only included those with a measured spectroscopic redshift of $z > 1.4$ (for comparison with work from the MOSDEF survey \citep{2015ApJ...815...98S}) and those objects that were assigned observations in {\it both} the prism and medium-resolution gratings. Additionally, we removed any objects whose observations were impacted by shorts in the MSA, most notably for many objects in GOODS-S observed under PID 1180. We additionally removed any objects that fell outside of the JWST/NIRCam coverage, lacked $>$$5\sigma$ detections in at least 3 photometric bands at $>$$0.7\rm\ \mu m$ from JWST or HST, did not have any robust rest-optical ($\lambda_{\rm rest} > 4000$\AA) or rest-UV ($\rm 1250\AA <\lambda_{\rm rest} < 2600\AA$) photometric observations, did not have a robust flux calibration solution (see Section \ref{sec:fluxcal} and Appendix \ref{sec:flux_cal_app}), or had poor-quality SED fits. We found that poor-quality SED fits fell into two categories: those with LRD-like SEDs \citep{2024ApJ...963..129M}, which {\sc prospector} struggled to fit accurately, and those for which the fitted SED model was clearly discrepant from the flux-calibrated prism spectra in the rest optical. In total, we identified 28 poorly fit SEDs. After applying these criteria, 1,164 of the JADES spectroscopic targets remained. After the addition of the AURORA objects with the same criteria applied, our total sample includes 1,204 objects at $z>1.4$. We show the distributions of stellar mass and UV magnitude (at 1600\AA) vs. redshift of objects in our sample in Figure \ref{fig:zdist}.

\subsection{Selection of Star-forming galaxies}\label{sec:selection}
From the parent sample of 1,204 spectroscopically confirmed galaxies at $z>1.4$, we analyze only a subset for the purpose of studying the SFMS. Firstly, we remove quiescent galaxies by excluding targets whose sSFRs, as calculated from their UV luminosity (see Section \ref{sec:uv_sfr}) are less than ${10^{-11}}\ \rm yr^{-1}$. The choice of how to select star-forming galaxies can affect the measured slope, normalization, and scatter of the SFMS \citep[see discussions in][]{2014ApJS..214...15S,2022ApJ...936..165L,2023MNRAS.519.1526P}. Besides a cut in sSFR, other common methods for distinguishing between star-forming and quiescent galaxies include those based on rest-frame colors (e.g. UVJ), sigma-clipping of the SFMS, and LBG selection (e.g.,
\citep{2004ApJ...617..746D,2009A&A...504..751S,2009ApJ...691.1879W,2010ApJ...709..644I,2011ApJ...735...86W,2012ApJ...754...83B,2017ApJ...847...76S,2023ApJ...943..166A}.

Studies such as
\citet{2022ApJ...936..165L} have shown that choices in defining a star-forming galaxy, whether it be via a sSFR cut, a UVJ selection \citep[e.g.,][]{2009ApJ...691.1879W,2011ApJ...735...86W}, or various other methods, can introduce systematic uncertainties at the level of 0.2 dex for the SFMS scatter and 0.5 dex for the SFMS normalization. However, these differences arise mainly for massive galaxies $\rm (\log{(M_*/M_\odot) \gtrsim 10.5})$. In addition, we require the detection of emission lines, which removes many quiescent galaxies to begin with. Thus, we choose to simply restrict our galaxy sample to objects for which $\rm \log{(sSFR/yr^{-1})} > -11$. Additionally, we restrict the sample to galaxies with $>$3$\sigma$ detections of at least two H\thinspace{\sc i} recombination lines for the purpose of dust correction, which removes 400 objects from the initial sample. We also remove objects whose SFRs are determined at $<$1$\sigma$ significance, where $\sigma$ is the confidence interval of the SFR evaluated in linear, rather than log space, cutting an additional 86 objects. We note that the removal of these objects does not significantly alter the final conclusions that we present. We further remove 6 galaxies with $\beta < -3.5$, where $\beta$ is the power law UV continuum shape following the form $f_\lambda \propto \lambda^\beta$ (see Section \ref{sec:uv_sfr} for discussion on calculating $\beta$), and we remove an additional 13 likely AGN-dominated galaxies that are flagged by broad Balmer-line components or whose \nii/H$\alpha$ line flux ratios exceed 0.5. In applying this AGN criterion, we have also required all spectra to cover the \nii\ and H$\alpha$ lines, removing 39 objects. Finally, upon re-measuring the redshifts of all objects during line fitting, we measured the redshift of one object to be slightly less than $z=1.4$, though its redshift as originally provided in the JADES spectroscopic catalog was slightly above $z=1.4$. As a result, this object was removed from the analysis. After applying these sample selection criteria, 659 galaxies remain, comprising the sample that is the subject of the main analysis in this work. We refer to this selected sample as the ``SFMS sample." 

\subsection{Comparison with photometric samples}\label{sec:sample_analysis}
\begin{figure*}
    \centering
    \includegraphics[width=\textwidth]{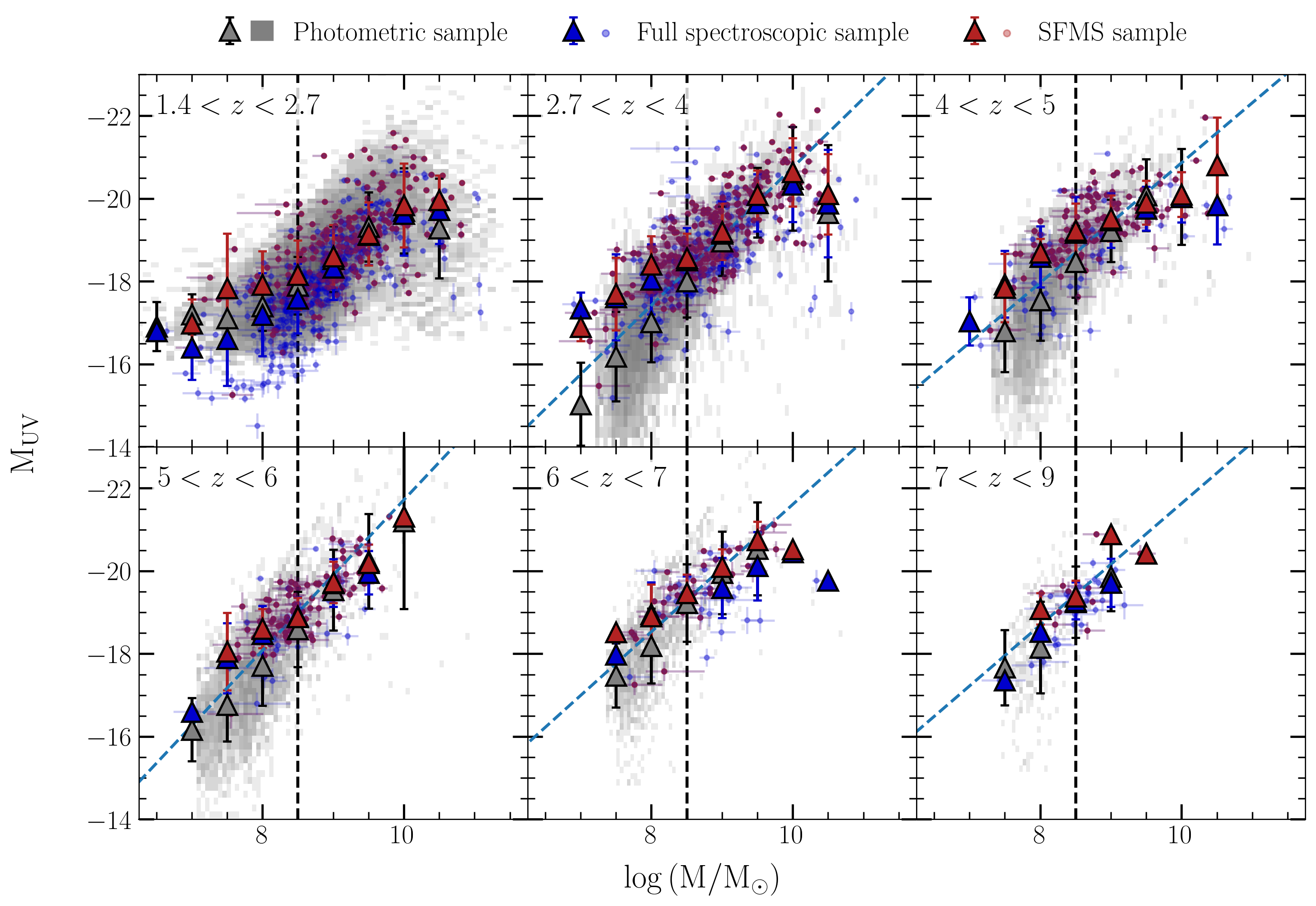}
    \caption{UV magnitude (at 1600\AA) vs. stellar mass for galaxy populations in the range $1.4 < z < 9$. The gray 2D histogram from $1.4 < z < 2.7$ represents measurements of galaxies in GOODS-N and GOODS-S from the 3D-HST program \citep{2014ApJS..214...24S}, while the gray 2D histograms at $z>2.7$ are from \citetalias{2024MNRAS.535.2998S}. The points in blue show the full spectroscopic sample that we analyze in this study, and the points in red show the galaxies which we use in our analysis of the SFMS. The blue, red, and gray triangles are the median $M_{UV}$ of the full spectroscopic sample, the SFMS sample, and the 3D-HST and \citetalias{2024MNRAS.535.2998S} samples, respectively, in bins of stellar mass. The error bars on the triangles show the 1$\sigma$ dispersion in $M_{UV}$ for each of the bin. We plot a vertical dashed black line at $10^{8.5}\rm\ M_\odot$ to visualize the representativeness limit. The blue diagonal dashed line shows the best-fit $M_{UV}$ vs. stellar mass relation from \citetalias{2024MNRAS.535.2998S}.}
    \label{fig:muv_mass}
\end{figure*}

Here, we present the combined JADES and AURORA spectroscopic sample in comparison with larger photometric samples. Given the complexity of the JADES NIRSpec target selection function, which follows a rank-ordered priority list for placing objects on the MSA \citep[see][]{2024A&A...690A.288B,2025ApJS..277....4D}, the representativeness of the NIRSpec sample is uncertain a priori. To better understand the population that the spectroscopic sample in this analysis represents, we choose to compare to larger, more complete photometric samples. Specifically, we compare our sample to HST data from the 3D-HST \citep{2012ApJS..200...13B,2014ApJS..214...24S} survey and JWST data from the JADES survey analyzed by \citet{2024MNRAS.535.2998S} (hereafter \citetalias{2024MNRAS.535.2998S}) as benchmarks.

The sample analyzed by \citetalias{2024MNRAS.535.2998S} consists of 14,652 galaxies from the JADES survey in GOODS-S, folding in data from the JEMS and FRESCO programs. The sample spans the redshift range $3.0<z<9.0$ and is 90\% mass complete down to a stellar mass of $\rm \log(M_*/M_\odot) \approx 7.5$. For the 3D-HST survey, the 90\% mass completeness limit is $10^{9.04}\rm\ M_\odot$ at $z<1.8$, and $10^{9.41}\rm\ M_\odot$ at $z<2.5$, based on stellar masses measured with {\sc FAST} \citep{Kriek_2009}. Due to the observed offset of $\sim$+0.2 dex for {\sc Prospector}-based stellar masses compared to {\sc FAST}-based masses observed in 3D-HST \citep{2019ApJ...877..140L} , the mass completeness limits from \citet{2014ApJ...789..164T} may be higher by $\sim$0.2 dex when considered in comparison with our spectroscopic sample. The mass limits from \citet{2014ApJ...789..164T}, however, represent a conservative estimate of the most shallow portions of the image mosaics, as the deep portions of the survey (such as HUDF) reach $\sim$1 mag deeper \citep{2014ApJ...789..164T,2014ApJS..214...24S}. Because both the 3D-HST and \citetalias{2024MNRAS.535.2998S} samples are complete down to relatively low masses, they serve as useful comparisons for the AURORA and JADES spectroscopic observations. We compare our sample with the aforementioned photometric samples in the $M_{UV}$ vs. $\log(\rm M_*/M_\odot)$ plane, since $M_{UV}$ is sensitive to both dust attenuation and SFR, making it a useful metric to explore how well our spectroscopic sample probes the SFMS. Comparing the distribution of our spectroscopic sample relative to the photometric sample allows us to ascertain how well the spectroscopic sample constrains the slope and normalization of the SFMS, the basis on which all scatter measurements are performed.

For galaxies in our $1.4 < z <  2.7$ bin, we compare to the 3D-HST sample, specifically the samples in GOODS-N and GOODS-S, in Figure \ref{fig:muv_mass}. The 3D-HST sample is shown as a gray 2D histogram, and the median $M_{UV}$ in bins of stellar mass is plotted as gray triangles, where the error bars are the 1$\sigma$ dispersion in $M_{UV}$. We also present our combined JADES and AURORA spectroscopic sample of 1,204 galaxies in blue points, and the blue triangles represent the stellar-mass-binned median $M_{UV}$. We refer to this sample as the ``Full spectroscopic sample." The red points and triangles represent the ``SFMS sample," the selection of which was described in Section \ref{sec:selection}. In the case of the $1.4 < z < 2.7$ bin, the distributions of galaxies in the photometric and SFMS samples trace each other closely in the mass range $7.5 < \rm \log(M_*/M_\odot) < 10.0$. Because the 3D-HST sample represents observations of varying depths, the close match in $M_{UV}$ vs. $\rm M_*$ with the SFMS sample suggests that a reasonable upper limit on the mass completeness of the SFMS sample is $\lesssim 10^{9}\rm\ M_\odot$. In Figure \ref{fig:muv_mass}, the slope of the $M_{UV}$ vs. $\rm M_*$ relation for the SFMS sample visibly flattens below $10^{8.5}\rm\ M_\odot$, suggesting that $10^{8.5}\rm\ M_\odot$ is a reasonable estimate of the mass completeness limit. In the mass range $10.0 < \rm \log(M_*/M_\odot) < 11.0$, the median $M_{UV}$ at fixed stellar mass is brighter in our spectroscopic sample than in the photometric sample. This $M_{UV}$ offset may indicate that our sample lacks massive, dusty star-forming galaxies and/or quiescent galaxies in this redshift range, which are fainter in $M_{UV}$. However, the mass-binned median $M_{UV}$ of the two samples differs by $<$0.5 mag in this mass range, suggesting that the spectroscopic sample is still fairly representative in the lowest redshift bin that we consider. This offset is present at $10^{10.5}\rm\ M_\odot$ in the $2.7<z<4$ bin as well, though to a lesser degree.

In the remaining redshift bins spanning the range $2.7 < z < 9.0$, we compare our spectroscopic sample to the \citetalias{2024MNRAS.535.2998S} sample. The differences between the \citetalias{2024MNRAS.535.2998S} sample and our spectroscopic sample are most pronounced at masses below $10^{8.5}\rm\ M_\odot$, where the median $M_{UV}$ of the spectroscopic sample at fixed stellar mass is consistently brighter than the \citetalias{2024MNRAS.535.2998S} sample by $\sim$0.5--1.5 mag. This offset towards higher $M_{UV}$ of the spectroscopic sample compared to the photometric sample indicates that, at masses below $10^{8.5}\rm\ M_\odot$, the spectroscopic sample is biased towards UV-bright galaxies and is missing the lower-SFR objects in the population. Above $10^{8.5}\rm\ M_\odot$, the sample distributions of $M_{UV}$ at fixed mass come into closer agreement, typically differing by $<$0.5 mag.

We also note the presence of an anomalous trend in the $4.0 < z < 5.0$ bin in which the median $M_{UV}$ at $\rm \log(M_*/M_\odot) \gtrsim 10$ of the spectroscopic samples lies at significantly {\it fainter} magnitudes than the best-fit trend line from \citetalias{2024MNRAS.535.2998S}. The median $M_{UV}$ of the spectroscopic samples in bins of stellar mass also follow a slope shallower than this trend line. This difference in slope and biased median $M_{UV}$ above $10^{10}\rm\ M_\odot$ likely arises due to very few UV-bright ($-22 < M_{UV} < -21.5$) objects being targeted in this mass range. This relative lack of UV-bright targets in this redshift bin has implications for deriving the properties of the SFMS, which we discuss in Section \ref{sec:results}. 

\section{Results} \label{sec:results}

\subsection{The Star-Forming Main Sequence}

\subsubsection{Calculating the H$\alpha$ SFR}\label{sec:neb_sfr}

We calculated the H$\alpha$-based SFR (SFR$_{\rm H\alpha}$) using a combination of at least two of the following six emission lines: Pa$\alpha$, Pa$\beta$, Pa$\gamma$, H$\alpha$, H$\beta$, and H$\gamma$. We only included objects with signal-to-noise ratio (SNR) measurements at $>$3$\sigma$ for at least two of the aforementioned lines. For certain portions of the spectra, notably in the wavelength ranges covered by the G140M and G395M gratings, we noticed several spurious features fitted as $>$3$\sigma$ detections, where the error spectrum is slightly underestimated for these features. Thus, for lines in these wavelength ranges, we use a SNR threshold of 5$\sigma$. Using the significantly detected H\thinspace{\sc i} line fluxes, we simultaneously fit for the $E(B-V)$ reddening and the dust-corrected SFR, assuming case B recombination \citep{2006agna.book.....O} at an electron temperature of 15,000 K and an electron density of 100 cm$^{-3}$. Using \texttt{PyNeb} \citep{2015A&A...573A..42L}, we calculated the intrinsic flux ratios of Pa$\alpha$, Pa$\beta$, Pa$\gamma$, H$\alpha$, and H$\gamma$ relative to H$\beta$ to be 0.30, 0.15, 0.09, 2.79, and 0.47, respectively. For all objects, we assumed a Milky Way \citep{1989ApJ...345..245C} dust attenuation law.

We constructed a likelihood function with $E(B-V)$ and $\rm \log{(SFR_{H\alpha})}$ as parameters, and we implemented a prior function restricting $E(B-V)$ to the range (0, 1). We sampled the resulting probability function using \texttt{emcee}, and we report the median value of the parameter samples as our fiducial $E(B-V)$ and $\rm \log{(SFR)}$ values, with the 16th and 84th percentiles of the sample distribution being the lower and upper bounds on the confidence interval for each parameter, respectively.

In order to convert from H$\alpha$ flux to $\rm \log{(SFR)}$, we first calculated the dust-corrected H$\alpha$ luminosity within the likelihood function. In cases where H$\alpha$ was not detected, but at least two other recombination lines had significant detections, we used the other recombination lines to infer a dust-corrected H$\alpha$ line flux, assuming the Case B recombination ratios described above. From the calculated or inferred H$\alpha$ luminosity, we then converted to SFR using the following equation:

\begin{equation}
    {\rm \log \left( \frac{SFR}{M_\odot\ yr^{-1}} \right)} = {\rm log \left( \frac{L_{H\alpha}}{erg\ s^{-1}} \right)} + C
\end{equation}

where $C$ is a conversion factor calculated using the Binary Population and Spectral Synthesis (BPASS) \citep{2017PASA...34...58E, 2018MNRAS.479...75S} binary models with an upper IMF limit of 100 $\rm M_\odot$ (see Section 2.3.3 of \citet{2022ApJ...926...31R} for more details). For the case of galaxies where we determined the fiducial SED fit to be the Calz+1.4Z$_\odot$ case (88 galaxies), we used a value of $C=-41.37$. For the case of galaxies where the SMC+0.28Z$_\odot$ assumption was the fiducial fit (571 galaxies), we used a value of $C=-41.59$, accounting for the dependence of the stellar metallicity on the conversion to SFR. 

\subsubsection{Calculating the UV SFR} \label{sec:uv_sfr}

We calculated the UV-based SFR (SFR$_{\rm UV}$) following the procedure outlined in \citet{2024ApJ...977..133C}. We first selected the photometric filter closest in wavelength to 1600 \AA\ in the rest frame with SNR $\geq 3$. Additionally, we calculated the UV slope $\beta$ from the photometry between 1250 \AA\ and 2600 \AA. For the star-forming spectroscopic sample, at a median redshift of $z_{median}=3.24$ and median UV magnitude of $M_{UV,median}=-19.1$, we calculate a median UV slope of $\beta = -1.77\pm0.02$, consistent with the findings presented in \citet{2014ApJ...793..115B}.

After calculating $\beta$, we calculated the attenuation at 1600 \AA\ ($A_{1600}$) assuming the following dust-law and metallicity-dependent relationships between $\beta$ and $A_{1600}$ calculated by \citet{2018ApJ...853...56R}:

\begin{equation}
    A_{1600} = \begin{cases}
        1.82\beta + 4.43, {\rm Calz+1.4Z_\odot}\\
        0.93\beta+2.52, {\rm SMC+0.28Z_\odot}
    \end{cases}
\end{equation}

From our derived value of $A_{1600}$, we then calculated the dust-corrected monochromatic UV luminosity at 1600 \AA\ ($\nu L_{\nu,1600}$), and calculated the UV-based SFR using the following equation:

\begin{equation}
    {\rm \log \left( \frac{SFR}{M_\odot\ yr^{-1}} \right)} = {\rm \log \left( \frac{\nu L_{\nu,1600}}{erg\ s^{-1}} \right)} + C
\end{equation}

where $C=-43.46$, which is the conversion factor calculated by \citet{2011ApJ...741..124H} and \citet{2011ApJ...737...67M}, adjusted to a \citet{2003PASP..115..763C} IMF. 

\subsubsection{Comparing the UV and H$\alpha$ SFRs with PROSPECTOR SFHs}

\begin{figure}[ht!]
    \centering
    \includegraphics[width=8.6cm]{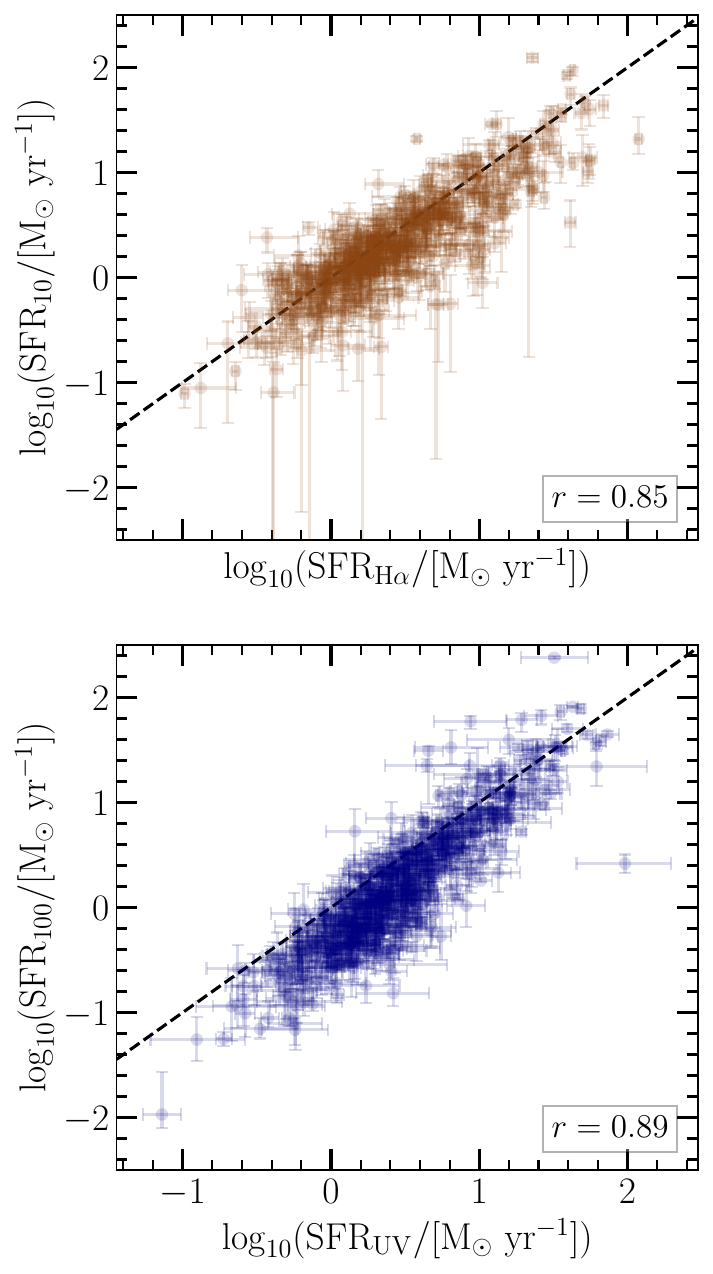}
    \caption{Comparison of the SFRs of 625 galaxies evaluated using SED-based vs. observational tracer-based estimates. {\it Top panel:} Comparison of the H$\alpha$-based SFR vs. the SFR from {\sc prospector} SED fitting averaged over the most recent 10 Myr. {\it Bottom panel:} Comparison of the UV-based SFR vs. the SFR from {\sc prospector} SED fitting averaged over the most recent 100 Myr. In both panels, the 1:1 line is shown as a black dashed line.}
    \label{fig:sfr_prospector}
\end{figure}

In Figure \ref{fig:sfr_prospector}, we compare the SFR values that we derive based on dust-corrected H$\alpha$ and UV luminosities (i.e., SFR estimates based on observational tracers) with those inferred from the non-parametric SFH {\sc prospector} fits (i.e., SED-based SFRs). The non-parametric SFHs are predicted to trace the true SFH well when compared to SFHs from simulations \citep[e.g.,][]{2020ApJ...904...33L,2024MNRAS.530L...7H}, thus, it is an intriguing exercise to compare the SED-based SFRs with the estimates from observational tracers. For the purpose of this comparison, we only evaluate star-forming galaxies from the JADES survey based on the selection criteria that we describe in Section \ref{sec:selection}.

Because the H$\alpha$ luminosity and the non-ionizing UV luminosity are sensitive to star formation on different time scales, SED-based SFHs averaged over the most recent 10 Myr and 100 Myr (SFR$_{10}$ and SFR$_{100}$, respectively) are often treated as analogous to emission-line-based and UV-based SFRs, respectively \citep[e.g.,][]{1998ARA&A..36..189K,2012ARA&A..50..531K,2025ApJ...979..193C,2025arXiv250804410S,2025arXiv250713160C,2025arXiv250804410S}. When considering these time scales as boxcar averages over the SFH, however, they may not exactly correspond to the H$\alpha$- and UV-based SFR indicators. \citet{2021MNRAS.501.4812F} suggest that the boxcar-averaged timescales that best match the H$\alpha$ and UV SFRs are closer to $\sim$5 Myr and $\sim$10 to $>$100 Myr, respectively, where the UV SFR time scale increases with SFH burstiness. Here, we briefly compare the observational tracer-based SFRs in our spectroscopic sample vs. the SED-based SFHs boxcar-averaged over the timescales of 10 Myr and 100 Myr commonly adopted in the literature. We calculate and report their Pearson correlation coefficients ($r$) in Table \ref{tab:sfr_correlation}. We include the boxcar-averaged time scales of 5 Myr and 50 Myr for comparison with the time scales suggested by \citet{2021MNRAS.501.4812F}.

In the top panel of Figure \ref{fig:sfr_prospector}, we show the comparison between SFR$_{10}$ and SFR$_{\rm H\alpha}$. The two SFR measurements trace each other well with a Pearson correlation coefficient of $r=0.85$. In the bottom panel of Figure \ref{fig:sfr_prospector}, we see that SFR$_{\rm UV}$ and SFR$_{100}$ are also well correlated, with a Pearson coefficient of $r=0.89$ and are slightly enhanced in SFR$_{\rm UV}$ relative to SFR$_{100}$ (a difference of $0.26\pm 0.29$, on average), which becomes more pronounced at SFR values below 10 M$_\odot\ \rm yr^{-1}$. This trend of enhanced UV-based SFR relative to SFR$_{100}$ at low SFRs is also seen in an analysis of the {\sc thesan-zoom} simulations \citep{2025arXiv250220437K} by \citet{2025arXiv250300106M} and in a sample from the JWST Emission Line Survey \citep[JELS;][]{2025MNRAS.541.1329D} by \citet{2025MNRAS.541.1348P}.

When we compare the correlation coefficients between different SFR indicators (Table \ref{tab:sfr_correlation}), we find that, in line with expectations, SFR$_{\rm H\alpha}$ is best correlated with SFR$_{10}$ as opposed to SFR$_{\rm 100}$, while SFR$_{\rm UV}$ is best correlated with SFR$_{100}$ as opposed to SFR$_{10}$. Additionally, we find minimal differences in correlation between SFR$_{5}$ vs. SFR$_{10}$ and SFR$_{50}$ vs. SFR$_{100}$ as they relate to SFR$_{\rm H\alpha}$ and SFR$_{\rm UV}$. This lack of difference in correlation between SFRs averaged over these particular time scales may simply reflect the coarse binning of the SFHs that we analyze, given that they represent only a factor-of-two difference in cosmic time, while the \citet{2021MNRAS.501.4812F} results are derived from much more finely time-resolved simulations. In any case, we do find that our data corroborate the picture that SFR$_{10}$ most closely traces the canonical short-time-scale SFR$_{\rm H\alpha}$, while SFR$_{100}$ most closely traces the canonical long-time-scale SFR$_{\rm UV}$ in our sample.

\begin{deluxetable}{c|cccccc}
\label{tab:sfr_correlation}
\caption{SFR correlation coefficients}
\tablehead{\colhead{} & \colhead{5} & \colhead{10} & \colhead{50} & \colhead{100} & \colhead{UV} & \colhead{H$\alpha$}}
\startdata
    5 & -- & -- & -- & -- & -- & --\\
    10 & 0.98 & -- & -- & -- & -- & --\\
    50 & 0.78 & 0.79 & -- & -- & -- & --\\
    100 & 0.74 & 0.74 & 0.97 & -- & -- & --\\
    UV & 0.83 & 0.84 & 0.89 & 0.89 & -- & --\\
    H$\alpha$ & 0.88 & 0.85 & 0.77 & 0.76 & 0.81 & --\\
\enddata
\tablecomments{The format of the first column and the column headers is such that each entry, $x$, (i.e., 5,10,50,100,UV,H$\alpha$) denotes the SFR$_x$ estimate.}
\end{deluxetable}

\subsubsection{Parametrizing the SFMS}

\begin{figure*}
    \centering
    \includegraphics[width=\textwidth]{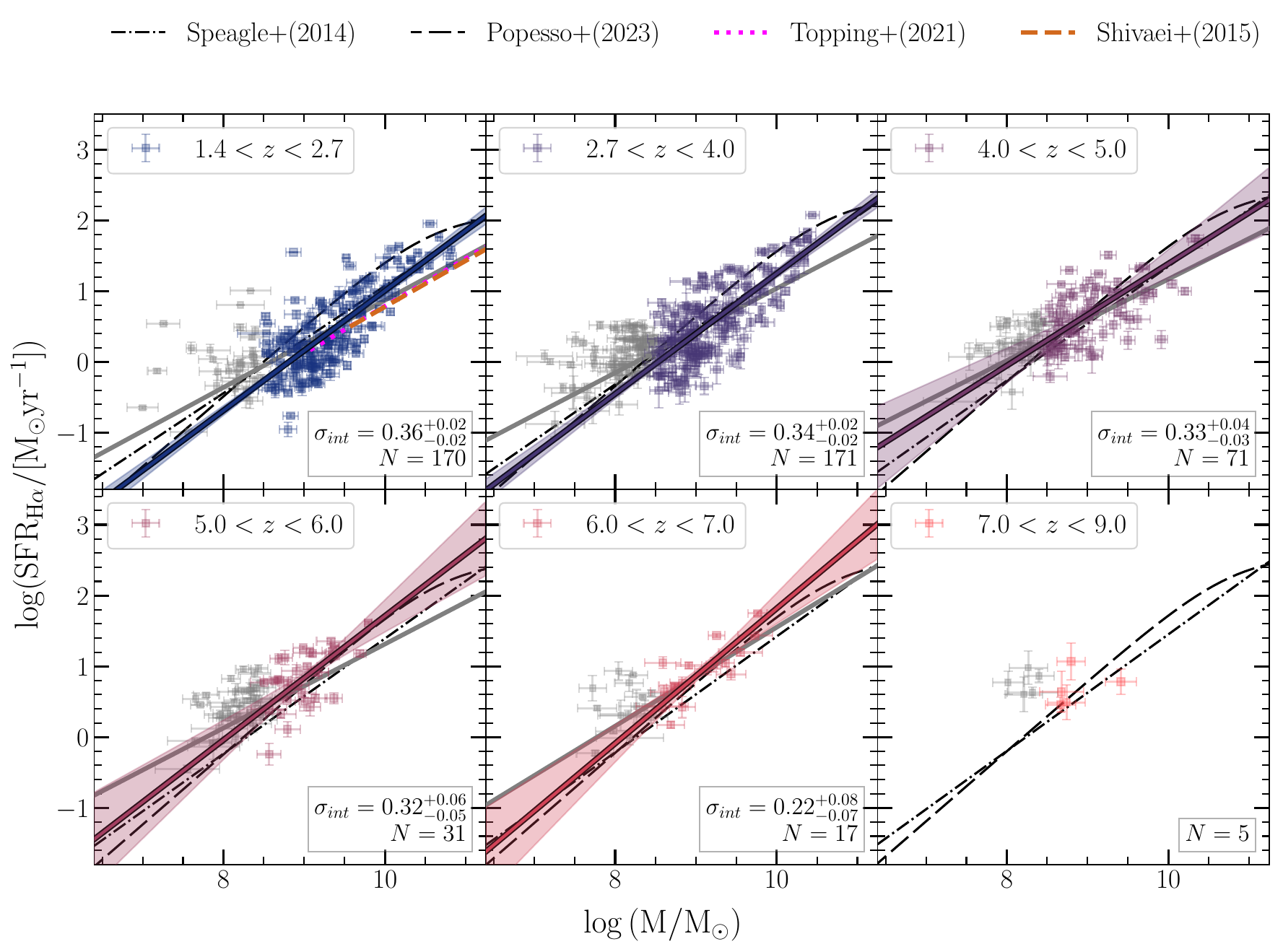}
    \caption{SFMS in bins of redshift, with SFR calculated from hydrogen recombination lines. The best linear fit to the full sample is shown as a solid gray line, while the best fit to galaxies above $10^{8.5}\rm\ M_\odot$ is shown as a solid colored line. The gray points have masses below $10^{8.5}\ \rm M_\odot$, while the colored points are above this mass threshold. SFMS measurements from the literature are also plotted. The intrinsic scatter, $\sigma_{\rm int}$, is shown in the bottom right corner of each plot, along with the number of objects in each redshift bin above $10^{8.5}\rm\ M_\odot$.}
    \label{fig:sfms_neb}
\end{figure*}

\begin{figure*}
    \centering
    \includegraphics[width=\textwidth]{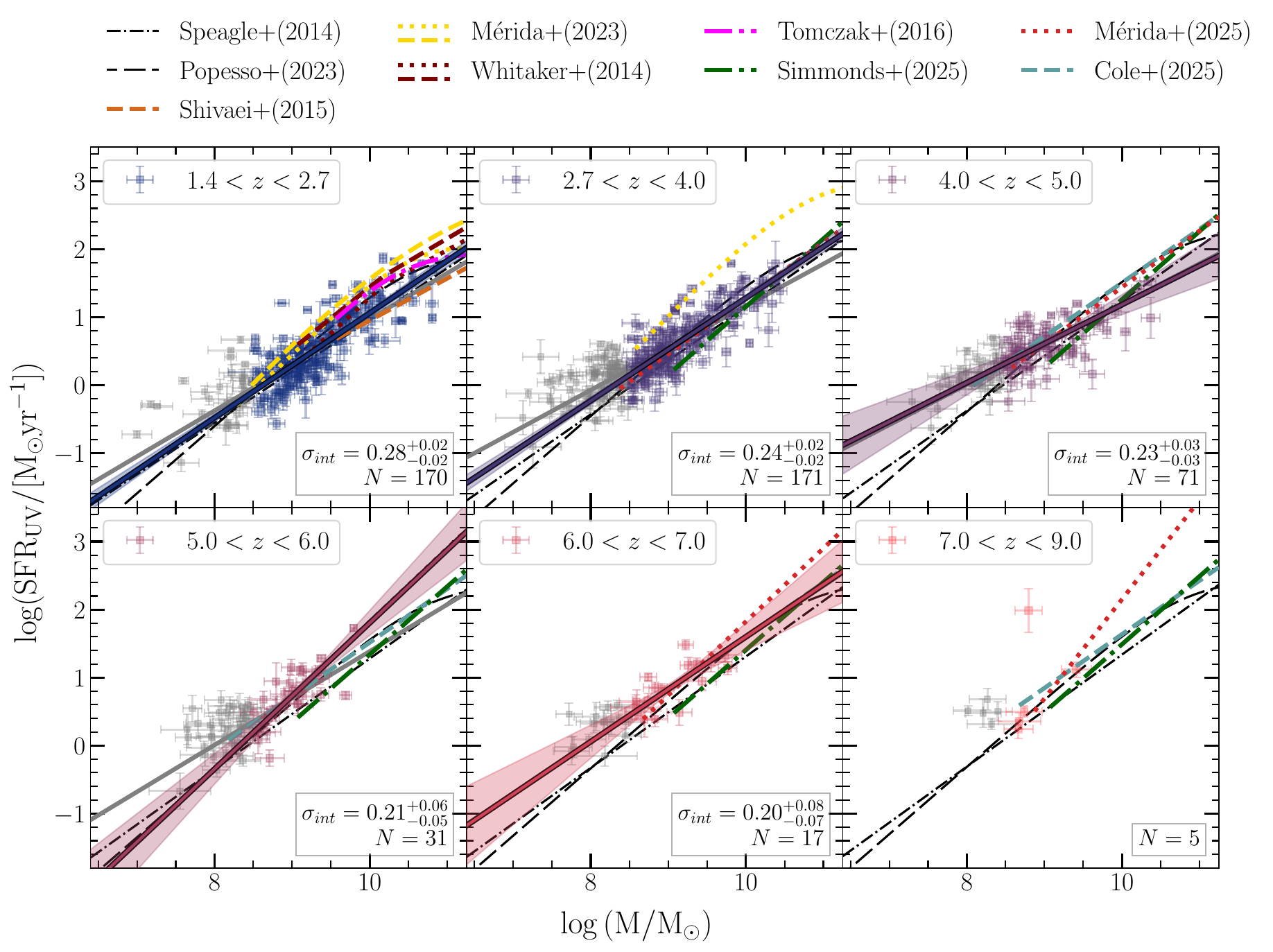}
    \caption{The SFMS in bins of redshift, with SFR calculated from the UV luminosity at 1600 \AA. The best linear fit to the full sample is shown as a solid gray line, while the best fit to galaxies above $10^{8.5}\rm\ M_\odot$ is shown as a solid colored line. The gray points have masses below $10^{8.5}\ \rm M_\odot$, while the colored points are above this mass threshold. SFMS measurements from the literature are also plotted. The intrinsic scatter, $\sigma_{\rm int}$, is shown in the bottom right corner of each plot, along with the number of objects in each redshift bin above $10^{8.5}\rm\ M_\odot$.}
    \label{fig:sfms_uv}
\end{figure*}

Throughout this study, we fit the SFMS with a single power law model. Several studies have noted that the SFMS slope exhibits a flattening at masses of $\rm \log(M_*/M_\odot) \gtrsim 10.5$, prompting some to fit the sequence with a broken power law \citep[e.g.,][]{2014ApJ...795..104W,2022ApJ...936..165L} or a functional form that flattens at high masses \citep[e.g.,][]{2014ApJ...795..104W,2015ApJ...801...80L,2020ApJ...899...58L, 2023ApJ...950..125M}. However, since the spectroscopic sample that we analyze in this study is almost entirely below this ``turnover" mass, we choose a single power law fit. We fit the data using the \texttt{linmix} Python package \citep{2007ApJ...665.1489K}, which accounts for measurement uncertainties in both the x- and y-axes, and calculates the intrinsic scatter about the regression line as a parameter. We fit the following linear model to the data:

\begin{equation}\label{eq:sfms}
\begin{split}
    \rm \log \left( \frac{SFR}{M_\odot \ yr^{-1}} \right) &= \alpha \log \left( \frac{M_*}{10^{9.11}\ M_\odot} \right) \\
    &+\beta_N + \mathcal{N}(0, \sigma_{int}^2)
\end{split}
\end{equation}

where $\alpha$ is the slope, $\beta_N$ is the intercept, and $\sigma_{\rm int}$ is the intrinsic, Gaussian random scatter about the relation, where $\sigma_{\rm int}^2$ = $\sigma^2 - \sigma_{\rm meas}^2$, $\sigma_{\rm meas}$ is the Gaussian scatter due to measurement error, and  $\sigma$ is the total scatter about the sequence. We normalize all of the masses by $\rm 10^{9.11}\ M_\odot$, the median stellar mass of the sample on the SFMS, as this normalization minimizes covariances between $\alpha$ and $\beta_N$ during fitting. We show the resulting SFMS in bins of redshift in Figures \ref{fig:sfms_neb} and \ref{fig:sfms_uv}. We do not fit the SFR vs. stellar mass relationship in the $7.0 < z < 9.0$ bin due to the small sample size, and due to the fact that we are unable to measure the \nii$\lambda 6585$/H$\alpha$ line ratio, which would allow us to rule out AGN-dominated objects consistently with the majority of the sample at lower redshift. We fit the SFMS for the full SFMS sample as well as restricting the fit to masses above $10^{8.5}\rm\ M_\odot$. We plot both of these fits as gray and colored lines, respectively in Figures \ref{fig:sfms_neb} and \ref{fig:sfms_uv}. Though we calculate and show the SFMS for the full spectroscopic sample, we only consider the fit to the SFMS sample for the remainder of the analysis.

\begin{deluxetable*}{c|ccccccc}
\label{tab:sfms_table}
\caption{Best-fit parameters for the SFMS shown in Figures \ref{fig:sfms_neb} and \ref{fig:sfms_uv}.}
\tablehead{
 \colhead{$z$ bin} & \colhead{$\alpha$} & \colhead{$\beta_N$} & \colhead{$\sigma_{\rm int}$} & \colhead{$\sigma_{\rm mass,1}$$\tablenotemark{\scriptsize a}$} & \colhead{$\sigma_{\rm mass,2}$$\tablenotemark{\scriptsize b}$} & \colhead{$\sigma_{\rm mass,3}$$\tablenotemark{\scriptsize c}$} & \colhead{$\rm N_{gal}$} \\ \hline \multicolumn{8}{c}{H$\alpha$ SFMS}}
 \startdata
\hline
$1.4 < z < 2.7$ & $0.85^{+0.05}_{-0.05}$ & $0.25^{+0.03}_{-0.03}$ & $0.36^{+0.02}_{-0.02}$ & $0.76^{+0.04}_{-0.03}$ & $0.39^{+0.01}_{-0.01}$ & $0.31^{+0.01}_{-0.02}$ & 170 \\
$2.7 < z < 4$ & $0.85^{+0.06}_{-0.06}$ & $0.49^{+0.03}_{-0.03}$ & $0.34^{+0.02}_{-0.02}$ & $0.58^{+0.02}_{-0.02}$ & $0.36^{+0.01}_{-0.01}$ & $0.31^{+0.02}_{-0.02}$ & 171 \\
$4 < z < 5^*$ & $0.72^{+0.21}_{-0.21}$ & $0.75^{+0.06}_{-0.06}$ & $0.33^{+0.04}_{-0.03}$ & $0.33^{+0.03}_{-0.03}$ & $0.33^{+0.02}_{-0.02}$ & $0.42^{+0.04}_{-0.04}$ & 71 \\
$5 < z < 6$ & $0.88^{+0.25}_{-0.24}$ & $0.93^{+0.07}_{-0.07}$ & $0.32^{+0.06}_{-0.05}$ & $0.47^{+0.03}_{-0.03}$ & $0.30^{+0.03}_{-0.03}$ & $0.18^{+0.08}_{-0.07}$ & 31 \\
$6 < z < 7$ & $0.95^{+0.23}_{-0.22}$ & $0.97^{+0.07}_{-0.07}$ & $0.22^{+0.08}_{-0.07}$ & $0.56^{+0.07}_{-0.06}$ & $0.20^{+0.05}_{-0.04}$ & $0.23^{+0.12}_{-0.22}$ & 17 \\
\hline 
\multicolumn{8}{c}{UV SFMS} \\ 
\hline
$1.4 < z < 2.7$ & $0.77^{+0.05}_{-0.05}$ & $0.37^{+0.03}_{-0.03}$ & $0.28^{+0.02}_{-0.02}$ & $0.47^{+0.03}_{-0.03}$ & $0.30^{+0.01}_{-0.01}$ & $0.33^{+0.02}_{-0.02}$ & 170 \\
$2.7 < z < 4$ & $0.76^{+0.04}_{-0.04}$ & $0.61^{+0.02}_{-0.02}$ & $0.24^{+0.02}_{-0.02}$ & $0.38^{+0.02}_{-0.02}$ & $0.27^{+0.01}_{-0.01}$ & $0.26^{+0.02}_{-0.02}$ & 171 \\
$4 < z < 5^*$ & $0.57^{+0.16}_{-0.15}$ & $0.67^{+0.05}_{-0.05}$ & $0.22^{+0.03}_{-0.03}$ & $0.24^{+0.03}_{-0.02}$ & $0.25^{+0.02}_{-0.02}$ & $0.25^{+0.08}_{-0.08}$ & 71 \\
$5 < z < 6$ & $1.08^{+0.20}_{-0.20}$ & $0.85^{+0.06}_{-0.05}$ & $0.21^{+0.05}_{-0.05}$ & $0.54^{+0.05}_{-0.05}$ & $0.17^{+0.04}_{-0.05}$ & $0.50^{+0.09}_{-0.07}$ & 31 \\
$6 < z < 7$ & $0.79^{+0.21}_{-0.20}$ & $0.91^{+0.07}_{-0.07}$ & $0.20^{+0.08}_{-0.07}$ & $0.27^{+0.06}_{-0.07}$ & $0.25^{+0.05}_{-0.04}$ & $0.22^{+0.10}_{-0.14}$ & 17 \\
\enddata
\tablenotetext{a}{Intrinsic scatter in the SFMS for galaxies at $\log(M_*/M_\odot) \leq 8.5$. Note that this mass range is below the mass representativeness limit of our sample}
\tablenotetext{b}{Intrinsic scatter in the SFMS for galaxies at $8.5 < \log(M_*/M_\odot) < 9.5$.}
\tablenotetext{c}{Intrinsic scatter in the SFMS for galaxies at $\log(M_*/M_\odot) \geq 9.5$.}
\tablenotetext{*}{We note that the sample in the $4<z<5$ bin lacks UV-bright objects at $\sim$$10^{10}\rm\ M_\odot$, biasing the SFMS slope to low values. We advise caution when interpreting results in this redshift bin.}\end{deluxetable*}

In addition to our measurements, we show SF sequences measured from different studies over similar ranges in galaxy stellar mass and redshift. In Figures \ref{fig:sfms_neb} and \ref{fig:sfms_uv}, we have shifted the literature curves to match our H$\alpha$- or UV luminosity-to-SFR conversion factors and adjusted to a \citet{2003PASP..115..763C} IMF. Since the SEDs of $\sim$$85\%$ of our sample are best fit by the SMC+0.28Z$_\odot$ dust law/metallicity combination, we shift the curves in Figure \ref{fig:sfms_neb} to match our $C=-41.59$ conversion factor. We find that the SFMS that we calculate broadly agrees with those presented in the literature, and we discuss the comparison to other works in more detail in Section \ref{sec:sfms_comparison}.

\subsubsection{The SFMS slope and normalization}

\begin{figure}
    \centering
    \includegraphics[width=8.3cm]{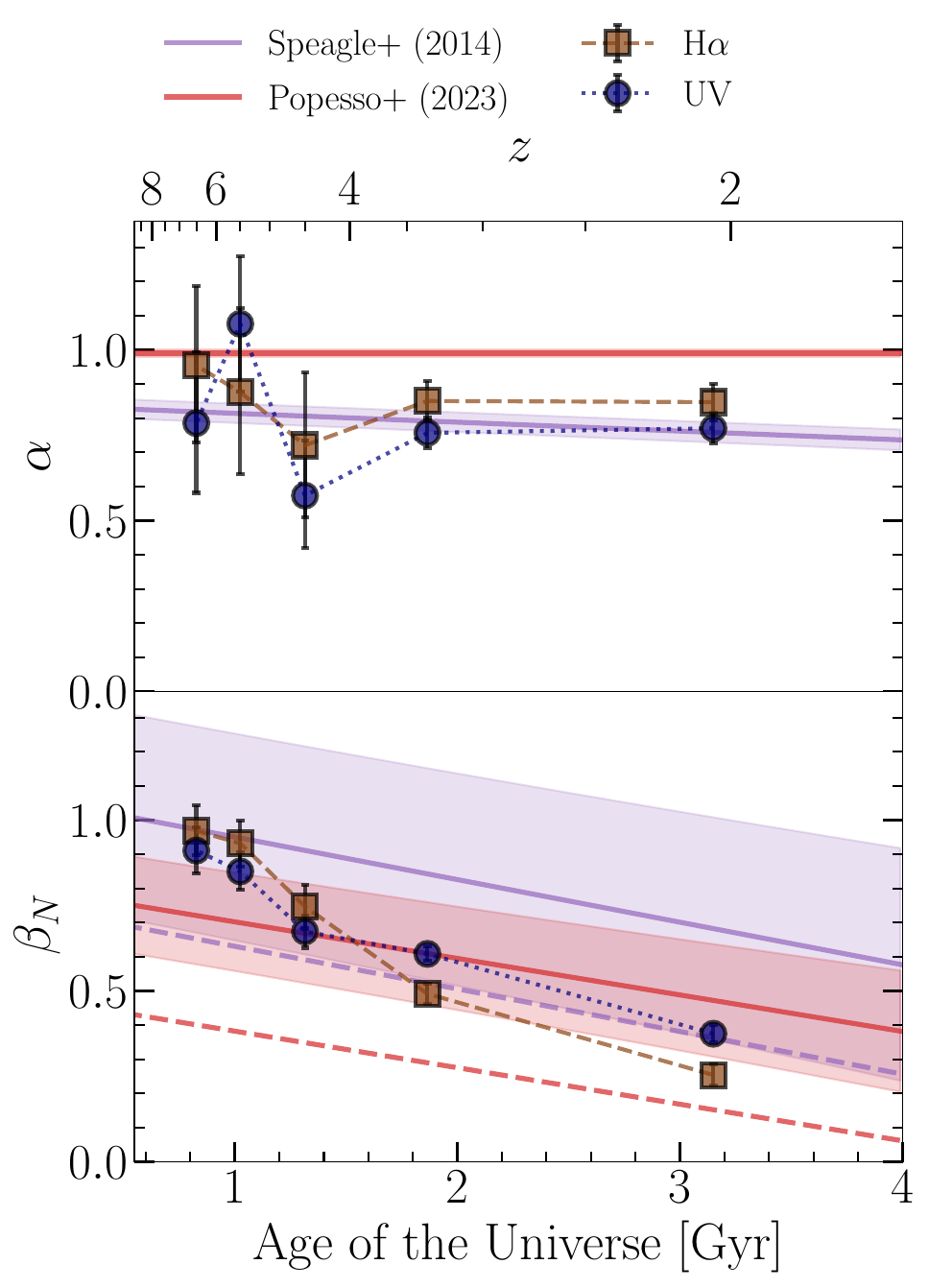}
    \caption{Slope $\alpha$ (top panel) and normalization $\beta_N$ (bottom panel) of the SFMS vs. cosmic time. The golden squares represent the SFMS parameters measured from the H$\alpha$-based SFMS, while the blue circles represent the SFMS parameters measured from the UV-based SFMS. The red and purple curves come from \citetalias{2023MNRAS.519.1526P} and \citetalias{2014ApJS..214...15S}, respectively. In the bottom panel, the dashed curves are the \citetalias{2023MNRAS.519.1526P} and \citetalias{2014ApJS..214...15S}, adjusted to match the $\rm L_{H\alpha}$ to SFR conversion that we use for low-metallicity galaxies.}
    \label{fig:alpha_beta_vs_z}
\end{figure}

We plot the values of $\alpha$ and $\beta_N$, defined in equation \ref{eq:sfms}, as a function of cosmic time in Figure \ref{fig:alpha_beta_vs_z}. Our dataset exhibits no strong relationship between the slope $\alpha$ and cosmic time when calculated with either the hydrogen recombination lines ($\alpha_{H\alpha}$) or the UV luminosity ($\alpha_{UV}$). We note the exception of an anomalous downturn in the slope for $4.0 < z < 5.0$, which is also visible in Figures \ref{fig:sfms_neb} and \ref{fig:sfms_uv}. We interpret this shallow slope as being due to a lack of UV-bright galaxies targeted in this redshift range. This lack of bright galaxies is also reflected in the fact that the median $M_{UV}$ of our spectroscopic sample at $\rm \log(M_*/M_\odot) = 9.5-10$ lies at fainter $M_{UV}$ than the best-fit relation of the \citetalias{2024MNRAS.535.2998S} photometric sample evaluated at the same stellar mass. When comparing $\alpha_{H\alpha}$ and $\alpha_{UV}$, the two values agree to within 1$\sigma$ across the full redshift range considered, though $\alpha_{H\alpha}$ is generally slightly steeper than $\alpha_{UV}$, apart from the $5<z<6$ bin. For reference, we also plot the cosmic-time-dependent slopes measured from the literature compilation studies by \citet{2014ApJS..214...15S} and \citet{2023MNRAS.519.1526P} (hereafter \citetalias{2014ApJS..214...15S} and \citetalias{2023MNRAS.519.1526P}, respectively). In the case of \citetalias{2023MNRAS.519.1526P}, we plot the fitted value of $\gamma$ that they present in Table 2 of their paper, where $\gamma$ represents the low-mass SFMS slope from their equation (11).

The SFMS slopes that we calculate are in agreement with some of the measurements in the literature that find a slope shallower than unity \citep[e.g.,][including \citealt{2022ApJ...936..165L} when fitting the mean of the SFMS]{2014ApJ...789...96A,2014ApJS..214...15S,2015ApJ...799..183S, 2015ApJ...815...98S,2021MNRAS.506.1237T,2025ApJ...979..193C}, while other studies find low-mass slopes more consistent with unity \citep[e.g.,][including \citealt{2022ApJ...936..165L} when fitting the ``ridge" of the SFMS]{2014ApJ...795..104W,2023ApJ...950..125M,2023MNRAS.519.1526P}.

Several works in the literature report a large range of slopes in the SFMS ranging from 0.4 to 1 \citep[see compilation studies by ][]{2014ApJS..214...15S,2023MNRAS.519.1526P}. Studies at $z \lesssim 2$ for which a substantial portion of the SFMS is fit at masses $\gtrsim$$10^{10}\rm \ M_\odot$ yield shallower SFMS slopes, while at higher redshifts, the low- and high-mass slopes come into closer agreement \citep[e.g.,][]{2014ApJ...795..104W,2015ApJ...801...80L,2022ApJ...936..165L}. Other observational effects that can influence the SFMS slope are numerous, including mass incompleteness \citep[e.g.,][]{2014ApJ...789...96A, 2022MNRAS.511.4464A,2025arXiv250804410S,2025arXiv250300106M}, target and star-forming galaxy selection methods \citepalias{2014ApJS..214...15S, 2023MNRAS.519.1526P}, dust correction methodologies \citep{2015ApJ...815...98S}, and mass-dependent metallicity effects \citep[e.g.,][]{2025arXiv250905403K}. We address mass incompleteness effects by fitting the SFMS to galaxies above $10^{8.5}\rm\ M_\odot$, and we see in Figures \ref{fig:sfms_neb} and \ref{fig:sfms_uv} that this sample cut results in steeper slopes than the full-sample SFMS, with the exception of the UV SFMS at $4<z<5$ and $6<z<7$ where the slope does not change.

We compare our measured slopes to those derived by \citetalias{2014ApJS..214...15S} and \citetalias{2023MNRAS.519.1526P} in Figure \ref{fig:alpha_beta_vs_z}. Similar to our findings, \citetalias{2014ApJS..214...15S} find little evolution in the SFMS slope in the redshift range probed by this study, remaining in the range 0.7--0.85. \citetalias{2023MNRAS.519.1526P} also find little evolution in the slope; however, they find values closer to unity.

We also plot the value of the SFMS normalization $\beta_N$ vs. cosmic time in Figure \ref{fig:alpha_beta_vs_z}. We illustrate the $\sim$0.3-dex adjustment in SFMS normalization due to the fact that we adopt a low-metallicity conversion from H$\alpha$ to SFR, plotting the corresponding \citetalias{2023MNRAS.519.1526P} and \citetalias{2014ApJS..214...15S} curves as dashed lines. We note that in our two lowest redshift bins spanning $1.4 < z <4$, we measure a higher value of $\beta_N$ in the UV-based SFMS ($\beta_{UV}$, not to be confused with the UV slope $\beta$) than in the H$\alpha$-based SFMS ($\beta_{H\alpha}$). At $z>4$, the two normalization estimates agree within 1$\sigma$. The phenomenon of a higher value of $\beta_{UV}$ than $\beta_{H\alpha}$ is a predicted consequence of bursty star formation \citep[][]{2019MNRAS.487.3845C}, as the FUV emission, tracing longer timescales of star formation, averages over short-timescale ($\lesssim 100$ Myr) dips in the SFR where the H$\alpha$ luminosity can briefly reach very low values.

\subsubsection{The SFMS intrinsic scatter}\label{sec:sfms_scatter}

In this section, we present the measurement-error-subtracted intrinsic scatter ($\sigma_{\rm int}$) about the SFMS across cosmic time. We first present the $\sigma_{\rm int}$ measurements from \texttt{linmix}, which incorporate data at stellar masses above $10^{8.5}\rm\ M_\odot$. These scatter measurements are presented for each redshift bin in Figures \ref{fig:sfms_neb} and \ref{fig:sfms_uv}, as well as in Table \ref{tab:sfms_table}. We find that the scatter in SFR$_{\rm H\alpha}$ vs. mass ($\sigma_{\rm int,H\alpha}$) is consistently larger than the corresponding scatter in SFR$_{\rm UV}$ vs. mass ($\sigma_{\rm int,UV}$) with the exception of the $6<z<7$ bin where the uncertainties on $\sigma_{\rm int}$ are large. We also observe a decrease in both $\sigma_{\rm int,UV}$ and $\sigma_{\rm int,H\alpha}$ with redshift, with a more pronounced evolution in $\sigma_{\rm int,H\alpha}$ than in $\sigma_{\rm int,UV}$.

To investigate how $\sigma_{\rm int}$ depends on stellar mass, we calculate the intrinsic scatter in three stellar mass bins: 

\begin{enumerate}
    \item $\rm \log(M_*/M_\odot) \leq 8.5$
    \item $\rm 8.5 < \log(M_*/M_\odot) < 9.5$
    \item $\rm \log(M_*/M_\odot) \geq 9.5$
\end{enumerate}

\begin{figure}
    \centering
    \includegraphics[width=8.3cm]{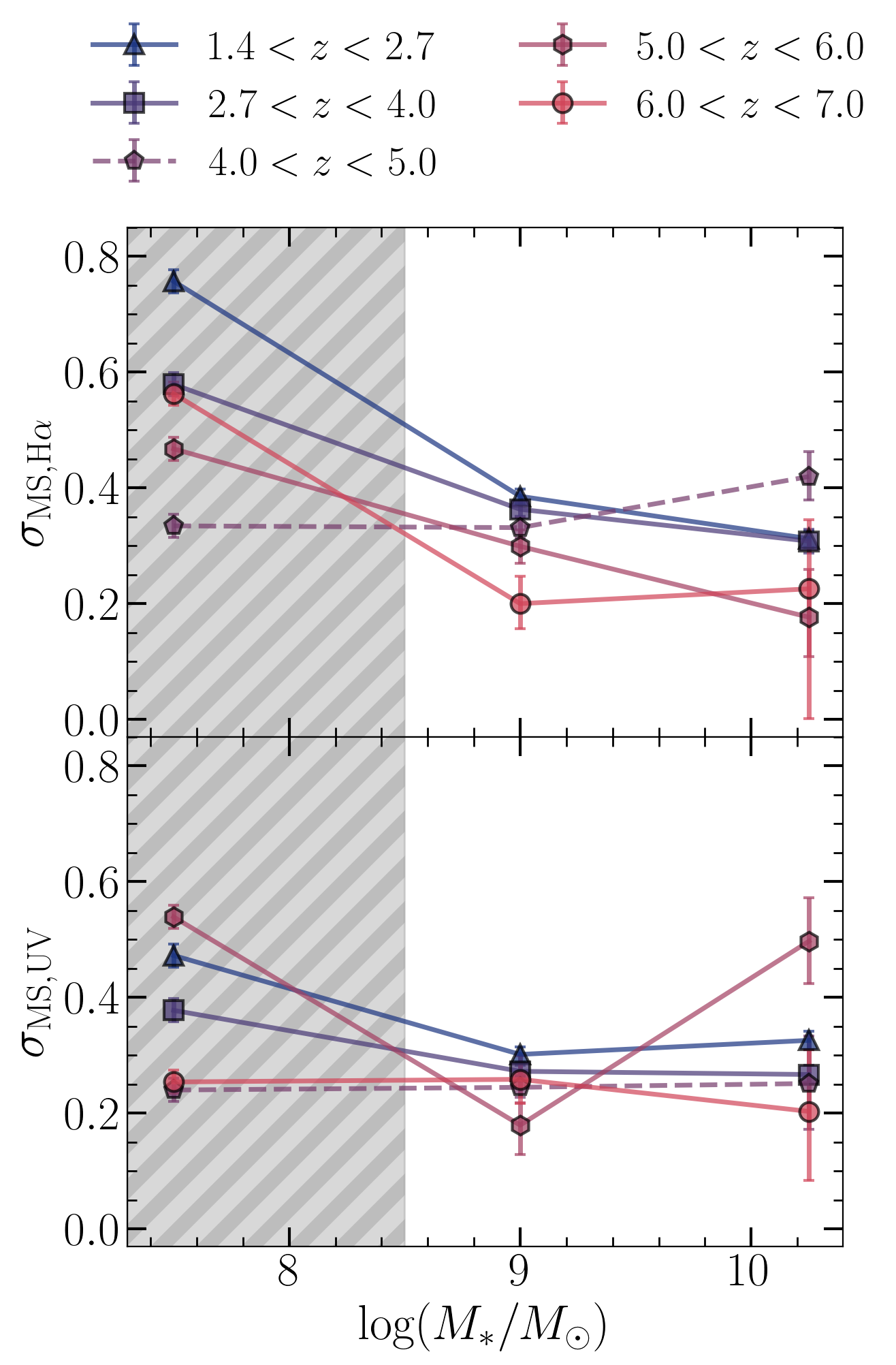}
    \caption{The measured intrinsic scatter about the SFMS as measured in the three mass bins presented in Table \ref{tab:sfms_table} ($\rm \sigma_{mass,1}$, $\rm \sigma_{mass,2}$, and $\rm \sigma_{mass,3}$) in each redshift bin. The $4<z<5$ bin is shown with a dashed curve as opposed to a solid curve, since the targets in this redshift range are biased toward low UV-magnitudes at high masses. The shaded region at M$_* \leq 10^{8.5}\rm\ M_\odot$ indicates the masses below which the sample is mass representative.}
    \label{fig:sfms_scatter}
\end{figure}

We refer to these scatter measurements as $\sigma_{\rm mass,1}$, $\sigma_{\rm mass,2}$, and $\sigma_{\rm mass,3}$, in the order listed. We calculate the mass-dependent scatter using an approach consistent with the procedure described in Section 3 of \citet{2024ApJ...977..133C}. We note that $\sigma_{\rm mass,1}$ encompasses galaxies below the mass completeness limit and is driven almost exclusively by galaxies that lie above the SFMS, so we caution against drawing firm conclusions from this measurement. The detection of additional fainter targets in this bin would provide better constraints on $\sigma_{\rm mass,1}$. We report $\sigma_{\rm mass,1}$, $\sigma_{\rm mass,2}$, and $\sigma_{\rm mass,3}$ in Table \ref{tab:sfms_table}, and we show them as a function of mass in Figure \ref{fig:sfms_scatter}. We point out that $\sigma_{\rm mass,3}>\sigma_{\rm mass,2}$ at $5<z<6$ for the UV-based SFMS, largely driven by two objects above $10^{9.5}\rm\ M_\odot$. Though these objects do not significantly bias the slope, they bias $\sigma_{\rm mass,3}$ high due to their very low inferred SFR uncertainties.

In the two lowest redshift bins spanning $1.4<z<4$ as well as $5<z<6$, we find that $\sigma_{\rm mass,2} > \sigma_{\rm mass,3}$ for the H$\alpha$-based SFMS. We refer the reader to Section \ref{sec:dust_uncertainties} for a discussion of the robustness of this result when adopting alternative nebular dust attenuation curves. Because $\sigma_{\rm mass,1}$ is based upon an extrapolation of the SFMS from masses above $10^{8.5}\rm\ M_\odot$, our reported value of $\sigma_{\rm mass,1}$ assumes symmetry about the SFMS below $10^{8.5}\rm\ M_\odot$. We therefore interpret this estimate of $\sigma_{\rm mass,1}$ as being consistent with an increasing $\sigma_{\rm int}$ with decreasing mass in the H$\alpha$-based SFMS. However, a representative sample of objects in this mass range is ultimately required to more robustly confirm whether $\sigma_{\rm int,H\alpha}$ continues to increase below $10^{8.5}\rm\ M_\odot$. The UV-based SFMS, in contrast, shows no trend of decreasing scatter with increasing stellar mass above $10^{8.5}\rm\ M_\odot$.

\subsection{The H$\alpha$/UV luminosity ratio}\label{sec:ha_uv_ratio}

\begin{figure*} 
    \centering
    \includegraphics[width=\linewidth]{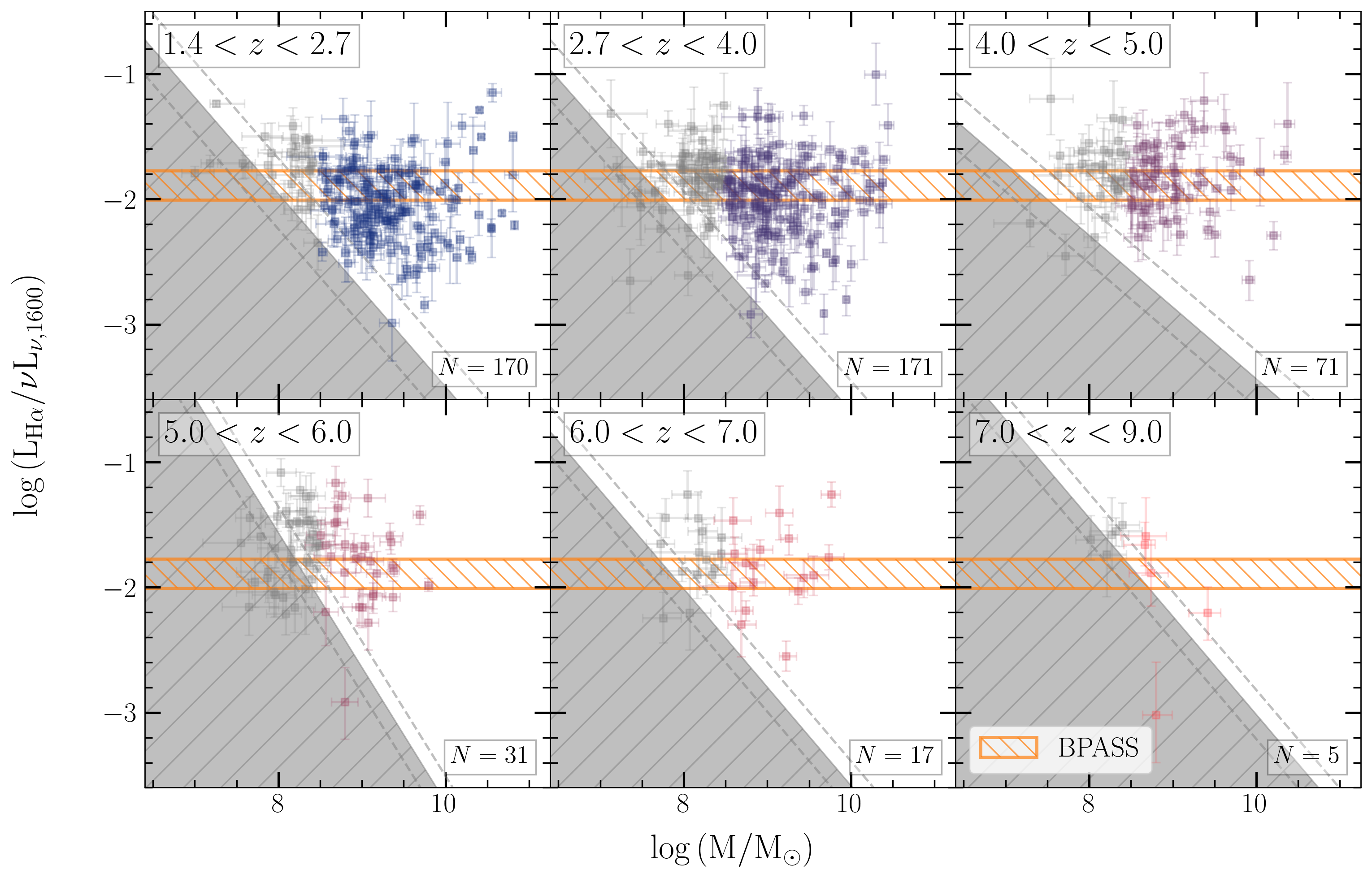}
    \caption{Ratio of dust-corrected H$\alpha$ luminosity to dust-corrected UV (1600\AA) luminosity ($\rm L_{H\alpha}/\nu L_{\nu,1600}$) as a function of redshift for the same spectroscopic sample as was analyzed for the SFMS. The coloring of the points corresponds to the redshift bin, while the gray points correspond to objects below 10$^{8.5}\ M_\odot$. The horizontal orange band represents the range of $\rm L_{H\alpha}/\nu L_{\nu,1600}$ values predicted with BPASS v2.2.1 models \citep{2017PASA...34...58E,2018MNRAS.479...75S} assuming a constant SFH and evaluated in the metallicity range $-2.15 < \log(Z/Z_\odot) < 0$. The gray hatched region shows the portion of the diagram that we are not sensitive to due to line sensitivity limits of the JADES observations. The dashed gray lines show the uncertainty on the boundary of the gray hatched region due to the $\sigma_{\rm int}$ measured from the UV-based SFMS.}
    \label{fig:ha_uv_ratio}
\end{figure*}

Here, we explore the ratio between the dust-corrected H$\alpha$ luminosity and the dust-corrected UV luminosity ($\rm L_{H\alpha}/\nu L_{\nu,1600}$). This quantity has been explored by several studies as a burstiness metric in addition to the analysis of the scatter about the SFMS \citep[e.g.,][]{1999MNRAS.306..843G, 2012ApJ...744...44W,2016ApJ...833...37G,2019ApJ...881...71E,2019ApJ...884..133F,2023ApJ...952..133M,2024MNRAS.52711372A,2025ApJ...987..189E}. Because the H$\alpha$ luminosity is a short-timescale tracer of the SFR, while the UV luminosity traces the SFR over longer timescales, the ratio of these two quantities can constrain the shape of a galaxy's recent SFH, distinguishing between rapidly rising and rapidly declining SFHs \citep[e.g.,][]{2015MNRAS.451..839D}. 

One strength of evaluating this metric is that it is a ratio of observable quantities, though the major systematic uncertainty involved in their measurement is the correction for dust attenuation. Systematic uncertainties associated with dust corrections have been identified as a substantial confounding factor in previous analyses of H$\alpha$/UV measurements, especially in cases where stellar attenuation is used to constrain the nebular attenuation \citep[e.g.,][]{2019ApJ...884..133F,2019ApJ...873...74B}. Even in the absence of dust, a significant drawback of this method for evaluating the burstiness of star formation is the degeneracy between different effects that influence the $\rm L_{H\alpha}/\nu L_{\nu,1600}$ ratio, with variations in ionizing spectrum shape and the presence of X-ray binaries being identified as possibly significant \citep[see ][]{2023MNRAS.526.1512R}. With this uncertainty in mind, we evaluate the $\rm L_{H\alpha}/\nu L_{\nu,1600}$ ratios in our sample to determine whether or not they are consistent with bursty star formation.

As part of this analysis, in addition to objects where H$\alpha$ is detected, we include objects that do not have coverage of the H$\alpha$ line at $z>6.7$, but instead have an H$\alpha$ luminosity inferred from the ratio of at least two dust-corrected H\thinspace{\sc i} lines, assuming Case B recombination as in Section \ref{sec:neb_sfr}. We display our measured $\rm L_{H\alpha}/\nu L_{\nu,1600}$ values in Figure \ref{fig:ha_uv_ratio} and plot measurements below $\rm 10^{8.5}\ M_\odot$ in gray to illustrate the limit below which the sample is most biased toward UV-bright galaxies.

In analogy with the analyses performed by \citet{2024MNRAS.52711372A} and \citet{2023ApJ...952..133M}, we compare our measured $\rm L_{H\alpha}/\nu L_{\nu,1600}$ values to predictions from BPASS v2.2.1 models \citep{2017PASA...34...58E,2018MNRAS.479...75S}. For a given set of SPS model assumptions, we calculate the value of $\rm L_{H\alpha}/\nu L_{\nu,1600}$ for a constant SFH and evaluate this ratio at $10^{8.5}\ \rm yr$ after the initial onset of star formation. We refer to this value as the ``equilibrium" $\rm L_{H\alpha}/\nu L_{\nu,1600}$ value for a given SPS model, and we calculate a range of equilibrium values, varying the stellar metallicity in the range $1\times10^{-4} < Z < 0.014$ ($-2.15 < \log(Z/Z_\odot) < 0$) and the IMF between \citet{2003PASP..115..763C}, \citet{2001MNRAS.322..231K}, and \citet{1955ApJ...121..161S} IMFs. The resulting equilibrium range spans the values of $-2.01 \leq \log(\rm L_{H\alpha}/\nu L_{\nu,1600}) \leq -1.68$ as shown in Figure \ref{fig:ha_uv_ratio} as a horizontal orange band. 

We also plot a gray shaded region in Figure \ref{fig:ha_uv_ratio}, as in \citet{2024ApJ...977..133C}, which illustrates, as a function of stellar mass, the $\rm L_{H\alpha}/\nu L_{\nu,1600}$ ratios that we do not expect to measure due to the limiting 3$\sigma$ NIRSpec line flux sensitivity for the JADES medium-tier observations. We describe our process for determining the limiting sensitivity curves in Appendix \ref{sec:sensitivity_curves_app}.

In Figure \ref{fig:ha_uv_ratio}, the effects of the limiting line flux sensitivity on the distribution of $\rm L_{H\alpha}/\nu L_{\nu,1600}$ vs. $\rm \log(M_*/M_\odot)$ are easy to visualize, since the lower left portions of the distributions in each redshift bin cut off parallel to the limiting sensitivity curve, with the exception of a handful of targets with deeper exposure times than the JADES medium-tier observations. This sensitivity limit is important to consider when analyzing the burstiness of star formation, since this sample contains few targets with very faint Balmer emission (11 objects with $\rm \log(SFR_{H\alpha})<-0.5$). Measuring faint Balmer-line targets is important because the H$\alpha$ line luminosity in bursty-SFH galaxies can reach zero shortly after the peak of an episode of star formation, while the UV light from B-type stars can linger for $\sim$100 Myr \citep{1999MNRAS.306..843G}. This rapid response of H$\alpha$ to changes in the SFH causes the intrinsic distribution of $\rm L_{H\alpha}/\nu L_{\nu,1600}$ to be skewed toward very low values in bursty galaxies \citep[e.g.,][]{2015MNRAS.451..839D,2019ApJ...873...74B}. \citet{2015MNRAS.451..839D} demonstrated that measuring $\rm L_{H\alpha}/\nu L_{\nu,1600}$ ratios of $\lesssim-2.5$ is required to fully capture the rapid quenching episodes of bursty SFHs in their simulations.

\begin{deluxetable*}{c|ccccccccc}
\label{tab:ha_uv_ratio}
\caption{Fraction of galaxies above and below equilibrium in the H$\alpha$/UV ratio.}
\tablehead{
 \colhead{$z$ bin} & \colhead{$f_{\rm eq}$} & \colhead{$f_{\rm eq,SFR}$} & \colhead{$f_{\rm above}$} & \colhead{$f_{\rm below}$} & \colhead{$\rm \log\left(\frac{L_{H\alpha}}{\nu L_{\nu,1600}} \right)_{med}$} & \colhead{$\rm \log\left(\frac{SFR_{10}}{SFR_{100}} \right)_{med}$} & \colhead{$\sigma_1$$\tablenotemark{\scriptsize a}$} & \colhead{$\sigma_2$$\tablenotemark{\scriptsize b}$} & \colhead{$\sigma_3$$\tablenotemark{\scriptsize c}$} }
\startdata
\hline
$1.4 < z < 2.7$ & 0.53 & 0.40 & 0.09 & 0.38 & $-2.04_{-0.02}^{+0.02}$ & $-0.12_{-0.03}^{+0.02}$ & $0.37^{+0.02}_{-0.02}$ & $0.25^{+0.01}_{-0.01}$ & $0.37^{+0.02}_{-0.01}$ \\
$2.7 < z < 4.0$ & 0.57 & 0.52 & 0.13 & 0.30 & $-1.98_{-0.02}^{+0.02}$ & $0.05_{-0.02}^{+0.02}$ & $0.26^{+0.02}_{-0.02}$ & $0.31^{+0.01}_{-0.01}$ & $0.35^{+0.03}_{-0.03}$ \\
$4.0 < z < 5.0$ & 0.62 & 0.52 & 0.27 & 0.11 & $-1.82_{-0.03}^{+0.03}$ & $0.17_{-0.04}^{+0.04}$ & $0.23^{+0.03}_{-0.04}$ & $0.26^{+0.02}_{-0.02}$ & $0.35^{+0.06}_{-0.05}$ \\
$5.0 < z < 6.0$ & 0.52 & 0.48 & 0.35 & 0.13 & $-1.77_{-0.04}^{+0.04}$ & $0.20_{-0.06}^{+0.06}$ & $0.30^{+0.03}_{-0.03}$ & $0.30^{+0.04}_{-0.04}$ & $0.27^{+0.05}_{-0.04}$ \\
$6.0 < z < 7.0$ & 0.59 & 0.59 & 0.24 & 0.18 & $-1.84_{-0.06}^{+0.06}$ & $0.27_{-0.10}^{+0.08}$ & $0.26^{+0.06}_{-0.06}$ & $0.29^{+0.04}_{-0.04}$ & $0.35^{+0.08}_{-0.07}$ \\
$7.0 < z < 9.0$ & 0.80 & 0.80 & 0.00 & 0.20 & $-1.92_{-0.17}^{+0.16}$ & $0.20_{-0.16}^{+0.18}$ & $0.29^{+0.11}_{-0.10}$ & $0.42^{+0.22}_{-0.19}$ & $0.30^{+0.00}_{-0.07}$ \\
\hline
\enddata
\tablenotetext{a}{Error-subtracted scatter in the $\rm L_{H\alpha}/\nu L_{\nu,1600}$ ratios about the sample median for galaxies at $\log(M_*/M_\odot) \leq 8.5$. Note that this mass range is below the mass representativeness limit of our sample}
\tablenotetext{b}{Error-subtracted scatter in the $\rm L_{H\alpha}/\nu L_{\nu,1600}$ ratios about the sample median for galaxies at $8.5 < \log(M_*/M_\odot) < 9.5$.}
\tablenotetext{c}{Error-subtracted scatter in the $\rm L_{H\alpha}/\nu L_{\nu,1600}$ ratios about the sample median for galaxies at $\log(M_*/M_\odot) \geq 9.5$.}
\end{deluxetable*}

With the sensitivity limits of the sample in mind, we quantify the burstiness of SFHs as probed by the $\rm L_{H\alpha}/\nu L_{\nu,1600}$ ratio using the metrics presented in Table \ref{tab:ha_uv_ratio}. The first set of metrics that we use describes the fraction of galaxies above, below, and within 1$\sigma$ of the equilibrium $\rm L_{H\alpha}/\nu L_{\nu,1600}$ value in each redshift bin. We refer to these metrics as $f_{\rm above}$, $f_{\rm below}$, and $f_{\rm eq}$, respectively. In the case of these metrics, we only consider galaxies at stellar masses greater than $10^{8.5}\rm \ M_\odot$. In the range $1.4 < z < 7$, we find an ``equilibrium fraction" ($f_{\rm eq}$) of 52\%--62\%; however, a $\rm L_{H\alpha}/\nu L_{\nu,1600}$ ratio in the equilibrium range does not guarantee a continuous star formation history, since a galaxy may have a bursty SFH, and we happen to observe the galaxy in a transition from a star-forming to a quiescent phase, and vice versa. Thus, the number of galaxies in the equilibrium range represents an upper limit. 

To estimate the number of truly smoothly, constant star-forming galaxies in our sample, we calculate the number of equilibrium galaxies whose {\sc prospector}-based SFHs also exhibit smooth, constant star formation. We quantify smooth, constant star formation with the criterion $|\rm \log(SFR_{10}/SFR_{100})| < 0.5$ and report the fraction of galaxies that satisfy both this and the $f_{\rm eq}$ criterion under the column $f_{\rm eq, SFR}$ in Table \ref{tab:ha_uv_ratio}. Removing targets with $\rm L_{H\alpha}/\nu L_{\nu,1600}$ in the equilibrium range whose SED-based SFHs indicate recent rapid changes in their SFH reduces the fraction of smoothly, constant star-forming galaxies to the range 40\%--59\%. This fraction suggests that the majority of galaxies in our sample at $z<6$ are poorly explained by a smooth, constant SFH.

In addition to analyzing the fraction of galaxies that is consistent with constant star formation, we estimate the measurement-subtracted scatter about the median $\rm L_{H\alpha}/\nu L_{\nu,1600}$ ratio in each redshift bin. We measure this scatter as a function of stellar mass, with $\sigma_1$ in the table representing the scatter among galaxies at masses lower than $10^{8.5}\ \rm M_\odot$, $\sigma_2$ representing galaxies in the range $10^{8.5}-10^{9.5}\ \rm M_\odot$, and $\sigma_3$ representing galaxies at masses larger than $10^{9.5}\ \rm M_\odot$. We do not observe any strong, consistent trend of scatter in the $\rm L_{H\alpha}/\nu L_{\nu,1600}$ ratios with stellar mass. Though we do measure a slight mass-dependence of the scatter about the SFMS, the lack of a similar trend of the scatter in the $\rm L_{H\alpha}/\nu L_{\nu,1600}$ ratio distribution is partly to be expected. In the case of a rapidly rising SFR, for example, both the H$\alpha$ and the UV luminosity will be elevated, causing a galaxy to rise above the SFMS. However, because both of these SFR indicators are elevated, the increase in the $\rm L_{H\alpha}/\nu L_{\nu,1600}$ ratio becomes less pronounced. Thus, measuring the mass dependence of the scatter in $\rm L_{H\alpha}/\nu L_{\nu,1600}$ may not be as informative as the mass-dependent scatter in the SFMS, especially in the case where survey line flux limits restrict the detection of low-$\rm L_{H\alpha}/\nu L_{\nu,1600}$ objects in the sample.

\subsection{The sSFR over Cosmic Time}\label{sec:ssfr_results}

\begin{figure*}[ht!]
    \centering
    \includegraphics[width=\linewidth]{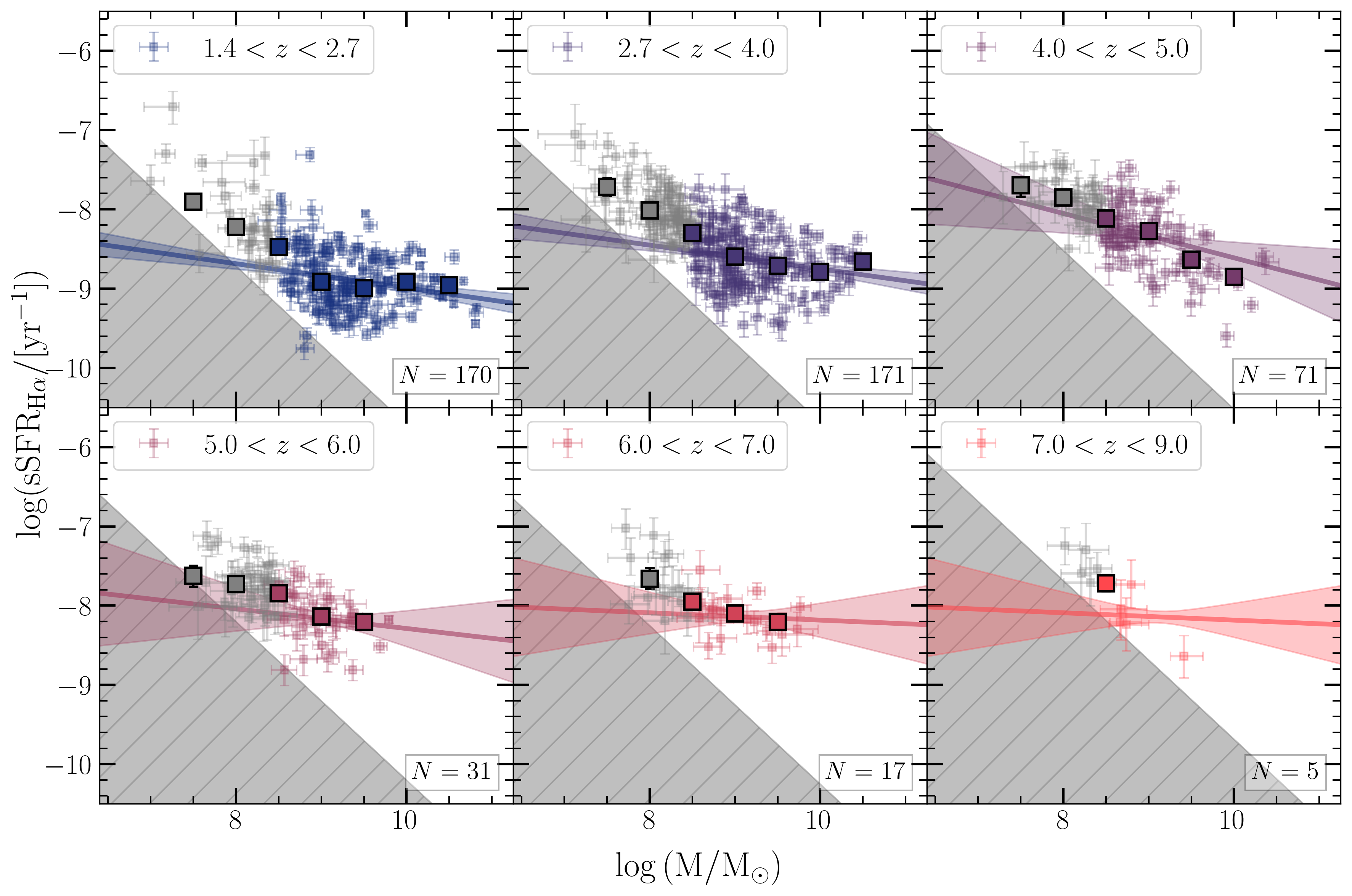}
    \caption{H$\alpha$-based specific SFR vs. stellar mass in bins of redshift. Binned median sSFRs are plotted as large squares, and the best fit to the SFMS with parameters from Table \ref{tab:sfms_table} are displayed as a solid colored line with a $1\sigma$ confidence interval shaded. The coloring of the points, solid lines, and shading correspond to the redshift bin. The points below $10^{8.5}\rm\ M_\odot$ are plotted in gray. The gray hatched shaded region shows the limited sensitivity region, where the sample is restricted by the depth of the NIRSpec observations. In the $7<z<9$ bin, since no fit to the SFMS was performed, we show the fit to the $6<z<7$ SFMS. $N$ denotes the number of galaxies in each redshift bin with a mass above $10^{8.5}\rm\ M_\odot$.}
    \label{fig:sSFR_nebular}
\end{figure*}

\begin{figure*}[ht!]
    \centering
    \includegraphics[width=\linewidth]{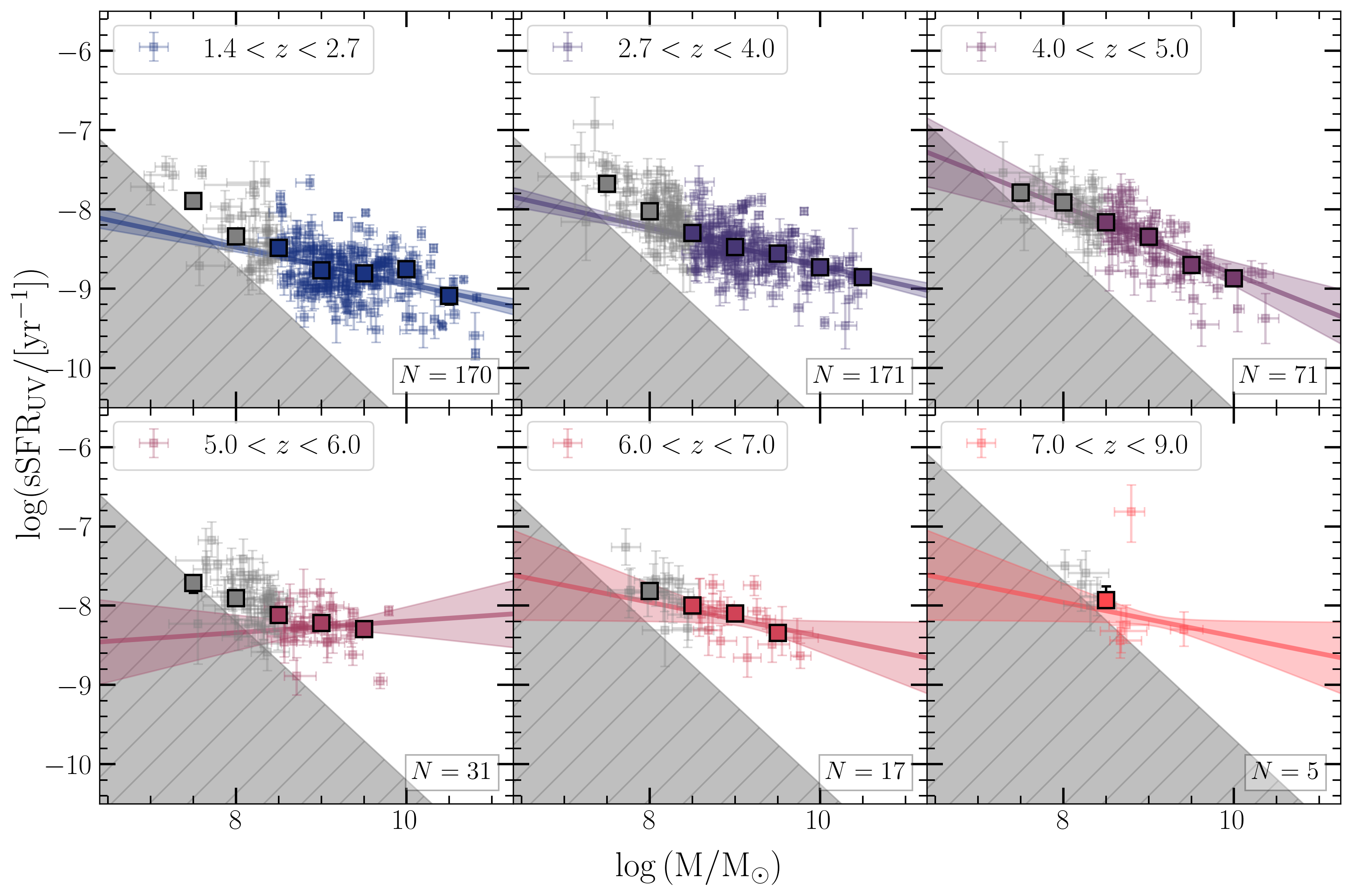}
    \caption{UV-based specific SFR vs. stellar mass in bins of redshift. Binned median sSFRs are plotted as large squares, and the best fit to the SFMS with parameters from Table \ref{tab:sfms_table} are displayed as a solid colored line with a $1\sigma$ confidence interval shaded. The coloring of the points, solid lines, and shading correspond to the redshift bin. The points below $10^{8.5}\rm\ M_\odot$ are plotted in gray. The gray shaded region shows the limited sensitivity region, where the sample is restricted by the depth of the NIRSpec observations. In the $7<z<9$ bin, since no fit to the SFMS was performed, we show the fit to the $6<z<7$ SFMS. $N$ denotes the number of galaxies in each redshift bin with a mass above $10^{8.5}\rm\ M_\odot$.}
    \label{fig:sSFR_uv}
\end{figure*}

In Figures \ref{fig:sSFR_nebular} and \ref{fig:sSFR_uv}, we show the H$\alpha$-based and UV-based sSFRs as a function of stellar mass and redshift, displaying points at masses below $10^{8.5}\rm \ M_\odot$ in gray. We also show the mass-binned medians as large squares in 0.5-dex stellar mass intervals, and we plot a gray hatched region to visualize how the restrictions imposed by the line flux sensitivity limits of the sample translate to the sSFR vs. stellar mass plane. Finally, we show the best-fit SFMS from Table \ref{tab:sfms_table} as a solid colored line. In the $7<z<9$ bin where no fit was performed, we show the fit to the data in the $6<z<7$ bin.

At all masses and redshifts, we find an anticorrelation between sSFRs vs. stellar masses for both the UV- and the H$\alpha$-based sSFRs. This anticorrelation partly reflects the fact that we measure a SFMS slope shallower than unity for both SFR indicators. At low masses, however, the anticorrelation is linked to the sensitivity limits of the survey. The effects of the limiting line sensitivity on our sample manifest similarly to the limiting sensitivity effects seen in the H$\alpha$/UV ratios in Figure \ref{fig:ha_uv_ratio}, whereby the sample distribution cutoff at low masses runs parallel to the sensitivity curve.

It is of interest to note that the distribution of sSFRs at masses below $10^{8.5}\rm \ M_\odot$ extends to much higher ($\sim$ 1 dex for sSFR$_{\rm H\alpha}$) sSFRs than for the galaxies at higher masses. This trend with stellar mass is also reflected in the median $\rm \log(sSFRs)$ and is a feature across all of the redshift bins up to $z=7$. Although much of the trend of high median sSFRs at low masses is likely driven by the sensitivity limits of the survey, an interpretation of the trend is that the distribution of $\rm \log(sSFR)$ widens with decreasing stellar mass. Assuming that the median value of $\rm \log_{10}(sSFR)$ vs. $\rm \log(M_*/M_\odot)$ follows the best-fit relation at all masses, the observed increase in median sSFR at low masses would imply that the observations presented in this paper represent the upper envelope of an intrinsically wide $\rm \log(sSFR)$ vs. $\rm \log(M_*/M_\odot)$ distribution at low masses. A widening of the sSFR distribution with decreasing mass would be consistent with bursty SFHs as a common mode of star formation among the high-$z$, low-mass galaxy population \citep[e.g.,][]{2018MNRAS.478.1694M}. To confirm the shape of the $\rm \log(sSFR)$ vs. $\rm \log(M_*/M_\odot)$ distribution, deeper spectroscopy of a representative sample of low-sSFR galaxies is needed.

\begin{figure}
    \centering
    \includegraphics[width=8.5cm]{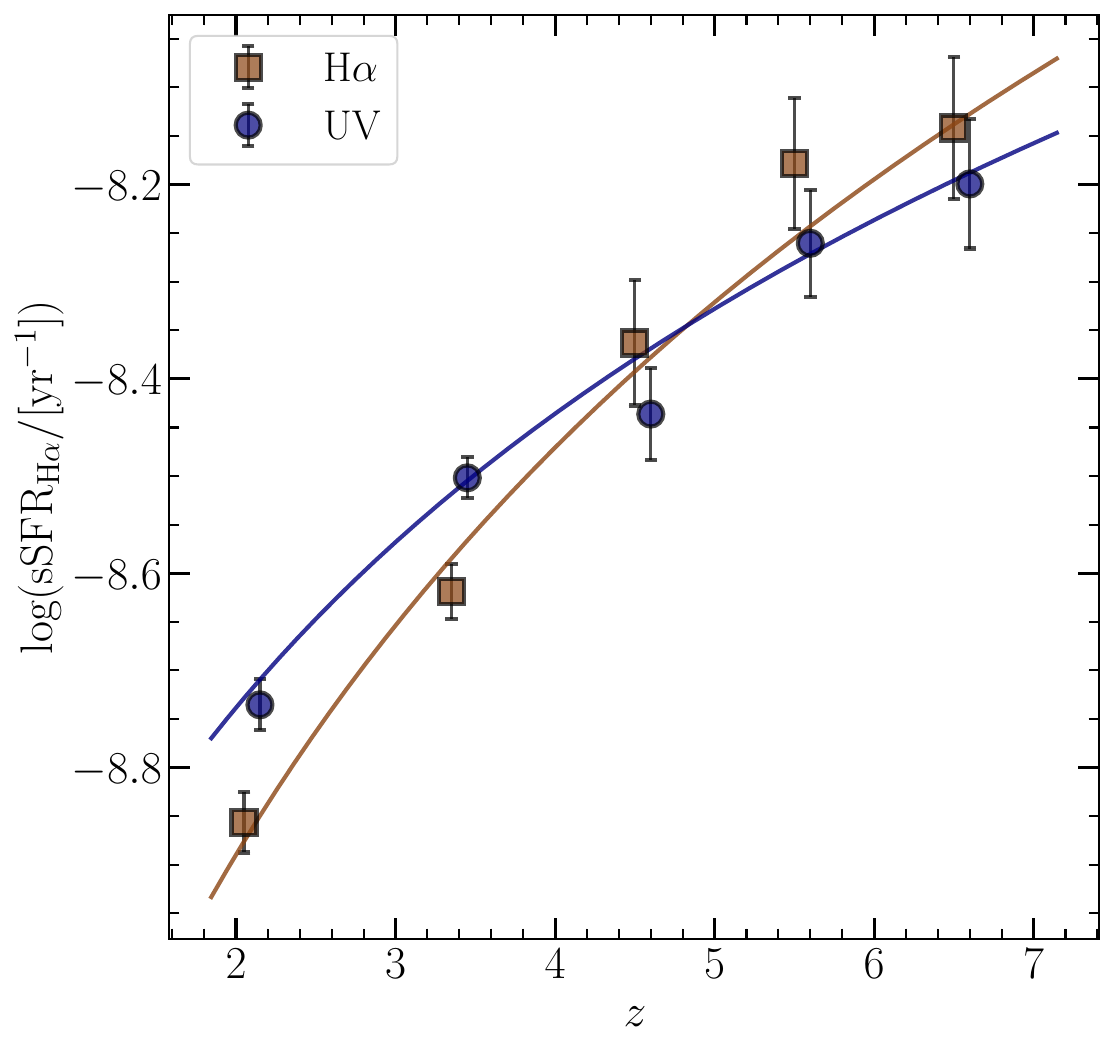}
    \caption{Median $\rm \log(sSFR)$ vs. redshift for the fitted SFMS at $10^{9.11}\rm\ M_\odot$. The colored curves correspond to the best-fit model from equation \ref{eq:ssfr_vs_z} to the data. The H$\alpha$ and UV points are all plotted at the same redshifts, with an offset of 0.1 for the purpose of visualization.}
    \label{fig:ssfr_fit}
\end{figure}

In Figure \ref{fig:ssfr_fit}, we show the redshift evolution of the median $\rm \log(sSFR)$ at $10^{9.11}\rm\ M_\odot$ for both the H$\alpha$- and the UV-based SFR indicators. We also fit these data using equation \ref{eq:ssfr_vs_z}:

\begin{equation}\label{eq:ssfr_vs_z}
    \log_{10} \left( {\rm \frac{sSFR}{yr^{-1}} }\right) = \gamma \log_{10}(1+ z) + \eta
\end{equation}

where $\gamma$ is the power-law slope in $(1+z)$, and $\eta$ is the normalization extrapolated to $z=0$.

The evolution of the median sSFR with redshift contains information similar to the SFMS normalization, $\beta_N$, normalized by the stellar mass. For this reason, it is not surprising that the normalization for the UV sSFR vs. redshift is higher than that of the H$\alpha$ sSFR vs. redshift curve at $1.4<z<4$, a trend which is consistent with bursty star formation \citep{2019MNRAS.487.3845C}. In examining the redshift evolution of the sSFR, we find power law slopes of $1.89^{+0.16}_{-0.15}$ and $1.36^{+0.13}_{-0.13}$ for the H$\alpha$- and UV-based sSFRs, respectively. These values are in rough agreement with recent studies in the literature, such as \citet{2025arXiv250804410S}, who find values between 1.06-2.30, depending on the SFR averaging timescale, with longer-timescale SFR measurements yielding smaller values of $\gamma$. We find smaller values for $\gamma$ than \citet{2022MNRAS.515.2951S}, however, who measure a value of $\gamma = 2.40^{+0.18}_{-0.18}$. In general, the evolution of our sSFR normalization is well described by a power-law evolution as $(1+z)^\gamma$, as has been found in the literature \citep{2013A&A...556A..55I,2014ApJS..214...15S,2014ApJ...795..104W,2015A&A...575A..74S,2020ApJ...899...58L,2021MNRAS.505..540T,2023MNRAS.519.1526P}.

\section{Discussion} \label{sec:discussion}

The core of this analysis is the comparison of inferred SFHs using SFR indicators sensitive to star formation on different time scales (namely, recombination lines vs. rest-UV emission) which have been independently corrected for dust attenuation. Several of the results presented in this work are consistent with bursty SFHs being common among the galaxy population at $1.4<z<7$, with a prevalence that is mass- and redshift-dependent.

Though the sample that we analyze is biased toward high-sSFR objects at masses below $10^{8.5}\rm \ M_\odot$, some information about SFHs at low masses can be inferred despite lacking Balmer-line detections of their low-sSFR counterparts. Through the remainder of this section, we describe how our SFMS measurements compare with other observational studies, discuss the ways in which our results are consistent with bursty SFHs, and describe how our results fit in the context of theoretical predictions from the literature.

\subsection{Comparison of the SFMS to Observational Works in the Literature}\label{sec:sfms_comparison}

In Figure \ref{fig:sfms_neb}, we compared our $1.4 < z < 2.7$ measurements to the SF sequences derived by \citet{2015ApJ...815...98S} and \citet{2021MNRAS.506.1237T}, both of which used dust-corrected measurements of Balmer lines from the MOSDEF survey \citep{2015ApJS..218...15K} to calculate SFRs. Though their mass completeness limits lie above those from the sample that we present, we find that their SF sequences are similar in normalization to the one that we present, though they measure shallower slopes ($0.65\pm0.08$ for \citet{2015ApJ...815...98S}, compared to our measured $0.85\pm 0.05$). These shallower slopes may reflect the line flux sensitivity of MOSDEF, as incompleteness at low masses typically flattens the SFMS, as demonstrated in Figures \ref{fig:sfms_neb} and \ref{fig:sfms_uv}, as well as \citet{2022MNRAS.511.4464A}.

In Figure \ref{fig:sfms_uv}, we showed our results in comparison with those found in the literature based on photometric observations. We again showed the \citetalias{2014ApJS..214...15S} and \citetalias{2023MNRAS.519.1526P} curves, and we found that they closely match our measured SFMS at all redshifts. Similar to the H$\alpha$-based SFR measurements from the MOSDEF survey, the UV-luminosity-based measurements presented by \citet{2015ApJ...815...98S} for the same MOSDEF targets closely match our results. Additionally, we compare to the studies by \citet{2014ApJ...795..104W} and \citet{2016ApJ...817..118T}, which use far IR emission in addition to FUV luminosity to account for dust-obscured star formation. Both studies find results consistent with our high stellar mass measurements in the lowest redshift bin. \citet{2023ApJ...950..125M} measured SED-based SFRs for galaxies down to $\rm \log(M_*/M_\odot) \approx8$ in the GOODS and CANDELS fields. Their derived SFMS lies $\sim$0.3 dex above our derived relation at $1.4 < z <2.7$, and the offset is more pronounced at $2.7 < z < 4$. A direct comparison between the normalization of these samples is complicated by the fact that their SFRs are measured through SED fitting, while ours are based on the dust-corrected UV luminosity. Additionally, systematics such as lack of accounting for outshining when deriving stellar masses in \citet{2023ApJ...950..125M} may contribute to the offset relative to our work \citep[see][for further discussion]{2025arXiv250922871M}. In any case, we include the \citet{2023ApJ...950..125M} results for comparison, since their sample is complete down to low stellar masses. A similarly mass-complete study was performed by \citet{2025ApJ...979..193C} using SED-derived SFRs from the JWST Cosmic Evolution Early Release Science (CEERS) survey \citep{2025ApJ...983L...4F}. The SFMS from this study agrees well with our findings across the redshift range $4.0<z<9.0$. We also compare with the recent studies by \citet{2025arXiv250804410S} and \citet{2025arXiv250922871M} and find similar agreement. Overall, our derived SFMS results agree well with those presented in the literature, where the small disagreements can largely be resolved when systematics and mass completeness limits are taken into account.

\subsection{Consistency with Burstiness}

\subsubsection{SFMS Scatter and Comparison to Theory}

In Section \ref{sec:sfms_scatter}, we explored how $\sigma_{\rm int}$ varies in bins of stellar mass and redshift. We found a decreasing $\sigma_{\rm int,H\alpha}$ with increasing stellar mass and no strong mass dependence of $\sigma_{\rm int,UV}$, while we found a trend of decreasing $\sigma_{\rm int,H\alpha}$ and $\sigma_{\rm int,UV}$ with redshift. Additionally, we found that in the range $1.4<z<4$, the SFMS normalization is higher for the UV-based SFMS than for the H$\alpha$-based SFMS. This difference in normalization, $\beta_N$ is consistent with expectations of a bursty SFH, whereby averaging over longer timescales results in a higher normalization with episodes of brief episodes of very low or zero star formation being smoothed out \citep[e.g.,][]{2019MNRAS.487.3845C,2019MNRAS.485.4817D,2020MNRAS.498..430I}.

Our finding of a higher $\sigma_{\rm int,H\alpha}$ with decreasing mass is qualitatively consistent with theoretical models and simulations that predict the SFHs of low-mass galaxies should be highly stochastic compared to higher-mass galaxies \citep[e.g.][]{2015MNRAS.451..839D,2017MNRAS.466...88S,2018MNRAS.478.1694M,2023MNRAS.525.2241H,2020MNRAS.497..698T,2024ApJ...975..192G}. Regarding the detailed predictions of the SFMS scatter, however, these theoretical works predict a wide range of scatter values for the mass regime considered in this study. Compared to the intrinsic SFMS scatter reported in this work, some theoretical models predict a lower scatter of $\sim$0.1--0.3 dex \citep[e.g.,][]{2015MNRAS.451..839D} and others predict a higher scatter of $\sim$0.4--0.7 dex \citep[e.g.,][]{2017MNRAS.466...88S,2023MNRAS.518..456D,2025arXiv250300106M}. Though the degree of scatter in theoretical works differs in detail from our findings, the qualitative trends are consistent with the predicted mass-dependent effects of the processes that modulate star formation. We also note that theoretical works that predict a small scatter may not strictly be inconsistent with our estimates, since systematic uncertainties on estimates of SFR and stellar mass may artificially boost our estimate of $\sigma_{\rm int}$. We discuss these possible effects in Section \ref{sec:uncertainties}.

The trend of decreasing $\sigma_{\rm int,neb}$ and $\sigma_{\rm int,UV}$ with redshift is intriguing, and may seem counterintuitive since short-term variability in SFRs is predicted to increase at high redshifts \citep[e.g.,][]{2017MNRAS.470.4698A,2018MNRAS.473.3717F,2020MNRAS.497..698T,2025arXiv250300106M}. However, in the redshift range $3<z<9$, \citet{2025arXiv250300106M} predict a decreasing scatter with increasing redshift and stellar mass, stating that processes that act on long timescales, such as galaxy environmental effects, contribute significantly to the SFMS scatter at low redshifts, while these effects are not as pronounced at higher redshifts, leading to an overall decreasing scatter with lookback time. If our measurement of decreasing scatter with redshift is robust, then it would support the aforementioned explanation. However, we note that sample sizes are lowest in the higher-redshift bins, and the limiting line flux sensitivity becomes increasingly restrictive with redshift. Future analyses with deeper spectroscopic observations of a larger ($\gtrsim$100) sample at $z > 5$ will provide more robust constraints on the evolution of $\sigma_{\rm int}$ with redshift.

In comparing our results to theoretical results in the literature, we find qualitative agreement in the trends of $\sigma_{\rm int}$ with redshift and mass, though in detail, our measurements differ quantitatively from theoretical works. We note the existence of systematic uncertainties that should be considered when interpreting our results, which we discuss further in Section \ref{sec:uncertainties}. To reach a strong consensus on the mass dependence of the scatter, future observational works will rely on large galaxy samples, representative down to $\sim$$10^{7.5}\rm\ - 10^8\  M_\odot$, representative galaxy samples complete down to $\sim$$10^{7.5}\rm\ - 10^8\  M_\odot$, similar to the mass completeness limits of photometric samples \citep[e.g., \citetalias{2024MNRAS.535.2998S};][]{2025ApJ...979..193C,2025arXiv250804410S,2025arXiv250922871M}.

\subsubsection{H$\alpha$/UV luminosity}

In Section \ref{sec:ha_uv_ratio}, we measured the H$\alpha$/UV ratios of the galaxies in our sample, finding that 41--60\% of galaxies are poorly described by constant SFHs. This finding is consistent with the picture that bursty SFHs are common at high redshift. We also note that in the range $4<z<7$, $f_{\rm above}$ is larger than in the range $1.4<z<4$, as can be seen in Table \ref{tab:ha_uv_ratio}. This trend is also reflected in the median $\rm L_{H\alpha}/\nu L_{\nu,1600}$ ratio for galaxies above $10^{8.5}\rm\ M_\odot$, labeled ``$\rm \log\left(\frac{L_{H\alpha}}{\nu L_{\nu,1600}} \right)_{med}$" in Table \ref{tab:ha_uv_ratio}. Conversely, $f_{\rm below}$ is larger in the range $1.4<z<4$ than it is at $4<z<7$. This trend suggests a possible evolution of increasing $\rm L_{H\alpha}/\nu L_{\nu,1600}$ ratios with redshift. There are several potential explanations for this trend. 

One interpretation may be that galaxies become increasingly metal-poor and $\alpha$-enhanced at fixed stellar mass with increasing redshift, a trend that has been observed in many studies \citep[e.g.,][]{2016ApJ...826..159S,2017ApJ...836..164S,2020MNRAS.499.1652T,2021ApJ...914...19S,2021MNRAS.505..903C,2023ApJ...957...81C}. These evolving chemical abundance patterns would naturally result in a harder ionizing spectrum, since at fixed metallicity, a higher $\alpha$/Fe ratio leads to a lower opacity for ionizing UV radiation in stellar atmospheres, boosting the H$\alpha$ luminosity relative to the non-ionizing UV continuum flux \citep[e.g.,][]{2025MNRAS.537.2433B}. The effects of variations in stellar abundance patterns are partly, albeit indirectly captured by the width of the orange band in Figure \ref{fig:ha_uv_ratio}, which is generated by varying the stellar metallicities of the BPASS models. Because the stellar ionizing UV spectra are impacted more strongly by the iron abundance than the abundance of $\alpha$ elements like oxygen, we can estimate the effects of $\alpha$-enhanced abundance patterns, assuming a fixed metallicity. The orange band in Figure \ref{fig:ha_uv_ratio} represents a variation in metallicity between 1\% vs. 100\% Z$_\odot$ resulting in a $\sim$0.2-dex difference in $\rm L_{H\alpha}/\nu L_{\nu,1600}$. This abundance-pattern evolution may partly explain the $\sim$0.27-dex increase in $\rm L_{H\alpha}/\nu L_{\nu,1600}$ from $1.4<z<2.7$ to $5<z<6$. 

\begin{figure*}
    \centering
    \includegraphics[width=\textwidth]{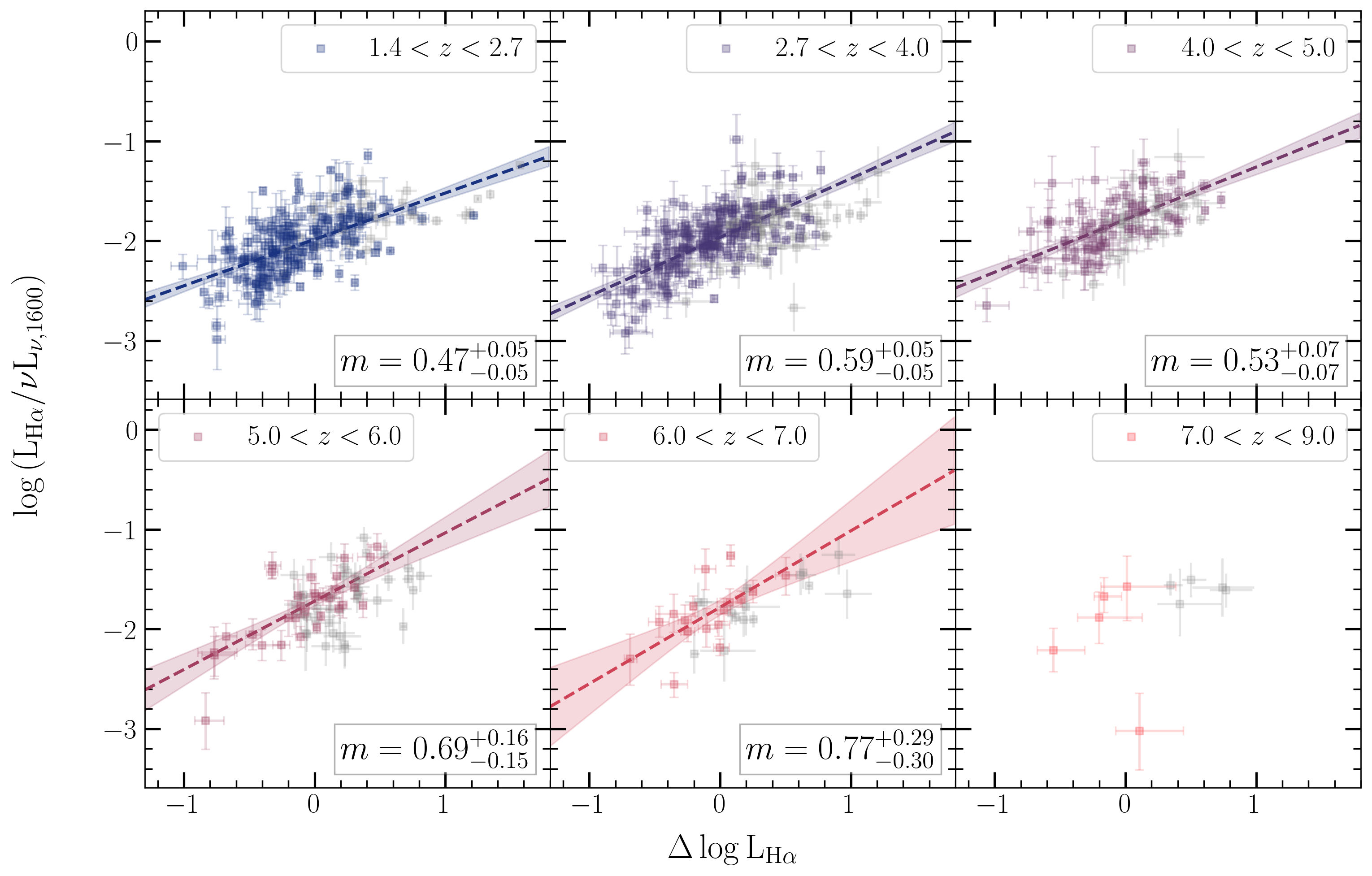}
    \caption{Distribution of $\rm L_{H\alpha}/\nu L_{\nu,1600}$ vs. SFMS offset ($\Delta\rm \log(L_{H\alpha})$) in each redshift bin. The colored dashed line represents a linear fit to the distributions in each redshift bin, and the slope $m$ is reported in the legend. The colors of the points correspond to the redshift bin. Objects below $10^{8.5}\rm\ M_\odot$ are plotted in gray and are excluded from the linear fit.}
    \label{fig:dlog_ha}
\end{figure*}

Alternatively, we briefly explore the possibility that the increasing $\rm L_{H\alpha}/\nu L_{\nu,1600}$ with redshift reflects a change in the behavior of galaxy SFHs at high redshift. For example, the elevated luminosity ratios may suggest that rising SFHs are more common with increasing redshift, consistent with expectations from theory \citep[e.g.,][]{2018ApJ...868...92T}. When examining the median {\sc prospector}-based $\rm SFR_{10}/SFR_{100}$ values, labeled under the column ``$\rm \log\left(\frac{SFR_{10}}{SFR_{100}} \right)_{med}$," we also see an increase in this ratio up to $z=7$. Another potential explanation of the evolution in $\rm L_{H\alpha}/\nu L_{\nu,1600}$ is that bursts of star formation occur more frequently at higher redshift. This increase in frequency would naturally result in a larger fraction of galaxies being measured in a high-$\rm L_{H\alpha}/\nu L_{\nu,1600}$ phase. To investigate this possibility, we examine the distribution of galaxies on the $\rm L_{H\alpha}/\nu L_{\nu,1600}$ vs. $\Delta\rm \log(L_{H\alpha})$ plane, where $\Delta\rm \log(L_{H\alpha})$ is the offset above or below the SFMS, in units of H$\alpha$ luminosity instead of SFR. As shown by \citet{2019ApJ...881...71E}, when modeling SFR bursts with an exponentially rising and falling component with a characteristic timescale, this distribution becomes steeper with decreasing burst timescale. We plot this distribution in Figure \ref{fig:dlog_ha}.

To examine how the timescale of SFR fluctuations evolves with cosmic time, we measure the slope, $m$, of the distribution in each of our redshift bins for galaxies with masses of $\rm \log(M_*/M_\odot) > 8.5$. We measure a trend of slightly increasing slopes with increasing redshift, consistent with the interpretation that the characteristic time scales for SFR fluctuations are shorter at higher redshift. If we assume that star-formation bursts occur back-to-back, then this increasing slope is consistent with an increased frequency of short-duration burst events in the high-redshift universe compared to the low-redshift universe. However, when interpreting this result, one must take into account that the sample is smaller at high redshfit, and there is a large uncertainty on the measured slope in the highest redshift bins. Ultimately, a larger sample with deeper spectroscopy will be necessary to test the robustness of this result.

Finally, because the $\rm L_{H\alpha}/\nu L_{\nu,1600}$ distribution is inherently skewed toward low values for a highly bursty galaxy population \citep[e.g.,][]{2015MNRAS.451..839D,2019ApJ...873...74B}, the sensitivity limits may artificially suppress $f_{\rm below}$. Though we attempt to address this effect by only considering galaxies above $10^{8.5}\ \rm M_\odot$, we note that line flux sensitivity limits may still prevent the detection of the most H$\alpha$-faint galaxies in the population, especially given that the sensitivity limits become more restrictive with increasing redshift. 

\subsection{Potential Sources of Uncertainty}\label{sec:uncertainties}

In this section, we discuss potential sources of systematic uncertainties on the derived properties that we present in this work. The first uncertainty comes from the fact that all of the galaxies in this analysis were observed through slits on the JWST/NIRSpec MSA. Because these slits are 0\secpoint2 wide, light from extended sources or slight misalignments of the target with the slit result in light losses that must be accounted for. Additionally, and especially for extended galaxies at lower redshift, the slit may only cover a particular portion of the target galaxy, meaning that the resulting spectrum contains contributions from select regions, rather than the entire galaxy. Scaling the spectra to match the photometry as we have done in this analysis can correct for wavelength-dependent slit losses to a large degree. However, if the spatial profile of the nebular emission does not closely match that of the stellar light, then the equivalent width of the emission lines will not be representative of the galaxy as a whole, introducing additional scatter in the SFR estimates. Since most of the galaxies in the JADES sample are relatively compact, we anticipate that slit losses will only marginally contribute to uncertainties in the SFRs. An analysis conducted by \citet{2024A&A...690A..64M} using NIRCam grism data showed that galaxy H$\alpha$ sizes are typically $1.18\pm0.03$ times larger than those of the rest-optical-continuum sizes. Since the rest-optical-continuum-dominated photometry is used to correct the flux calibration of the H$\alpha$ line, we approximate the uncertainty due to slit losses as the square of the ratio of the typical H$\alpha$ vs. optical continuum sizes. This exercise suggests an uncertainty in H$\alpha$ luminosity due to slit losses of $\sim 0.14$ dex. Added in quadrature with an intrinsic SFMS scatter of 0.2 dex, the slit losses would only contribute 0.04 dex to the total scatter. Thus, we do not expect slit losses to be a dominant source of scatter. In addition, we have verified that the removal of the extended sources discussed in Appendix \ref{sec:ext_objs} does not alter the conclusions regarding the mass dependence of the SFMS scatter. Future works comparing NIRSpec slit spectra to slitless spectra of the same objects will be valuable for constraining the level of uncertainty introduced by slit losses.

Assumptions inherent in the conversion between Balmer-line luminosity and SFR also may contribute to uncertainties, such as our assumption of either $\rm 0.28Z_\odot$ or $\rm 1.4Z_\odot$ stellar metallicity and an ionizing radiation escape fraction ($f_{esc}$) of 0\%. A recent study by \citet{2025arXiv250905403K} based on the SPHINX simulations \citep{2018MNRAS.479..994R,2023OJAp....6E..44K} showed that accounting for individual galaxy metallicities may reduce the SFMS scatter by $\sim$0.04 dex, and increase the SFMS slope by $\sim$0.08 when compared to assumptions based on a single stellar metallicity. An observational study by \citet{2025ApJ...984..188K} of galaxies in the KBSS survey \citep{2012ApJ...750...67R, 2014ApJ...795..165S} found that adopting a metallicity-dependent SFR conversion factor steepened the SFMS slope by 0.03 dex. We partly capture metallicity variations in the population by utilizing SFR conversion factors that are consistent with the best-fitting metallicity to the galaxy SEDs. We thus do not anticipate that variations in stellar metallicities will significantly alter the SFMS scatter measurements that we make. Regarding the value of $f_{esc}$, \citet{2021MNRAS.505.2447P} measure values of 5--10\% at $z\sim 3$, while studies of objects at $z>4$ with similar UV brightnesses to our SFMS sample ($M_{UV}\lesssim  -18$) infer values closer to 10--13\% \citep{2024A&A...685A...3M,2025arXiv250701096G}, with the majority of objects consistent with $f_{esc}\lesssim0.1$. Additionally, the galaxies with highest $f_{esc}$ tend to be bluer and fainter ($M_{UV}\sim-17.5$, $\beta\sim -3.0$) than the galaxies in our SFMS sample \citep[e.g.,][]{2022ApJ...941..153T,2023MNRAS.524.2312E}.

Finally, we highlight our assumption of a \citet{2003PASP..115..763C} IMF throughout this analysis, which may lead to differences in SFR and stellar mass estimates if the IMF is not universal \citep[e.g.,][]{2022MNRAS.514L...6P,2022MNRAS.510.5603K,2024ApJ...963...74W,2024ARA&A..62...63H,2024ApJ...969...95Y,2024MNRAS.529.3563T}. Though it is not currently feasible to directly measure the IMF in galaxies at high redshift, uncertainties in the form of the IMF will affect inferred SFRs and stellar masses at the 0.3--0.4 dex level \citep{2024ApJ...963...74W}.

\subsection{Uncertainties regarding the nebular dust attenuation law}\label{sec:dust_uncertainties}

An additional uncertainty that we highlight is our assumption of a \citet{1989ApJ...345..245C} law to correct the nebular emission lines for dust attenuation, as well as the use of either an SMC or \citet{2000ApJ...533..682C} dust law for the stellar light, as is standard practice \citep{1994ApJ...429..582C,2015ApJ...806..259R,2018ApJ...853...56R}. Variations of the dust attenuation curve in individual galaxies, for example, may introduce artificial scatter into the derived, dust-corrected SFRs. Dust curve deviations can arise due to varying dust-to-star geometry or dust grain properties such as chemical composition and size \citep[e.g.,][]{2020ARA&A..58..529S}. Analyses of deep spectra from the AURORA survey have indicated that a variety of nebular dust attenuation curves may be characteristic of the high redshift galaxy population. \citet{2025ApJ...989..209S} presented an extreme example of a galaxy from the AURORA survey at $z=4.41$ which has a derived nebular dust curve that deviates significantly from \citet{1989ApJ...345..245C}. A recent study by \citet{2026ApJ...999...15R} shows evidence of deviations in the nebular dust curves of 24 galaxies at $z=1.52-4.41$, attributing much of the differences to galaxy-to-galaxy variations in the dust covering fraction toward OB associations. Additionally, a recent analysis by \citet{2025arXiv250901795S} reveals that stellar dust attenuation curves become flatter with increasing redshift. Though the majority of the galaxies above $10^{8.5}\rm\ M_\odot$ in our sample have only moderate inferred dust obscuration, with median values of $A_V=0.14\pm0.00$, $A_{H\alpha}=0.43\pm0.01$, $A_{1600}=1.01\pm0.02$ (denoting the stellar attenuation at 5500\AA, the nebular attenuation at H$\alpha$, and the stellar attenuation at 1600\AA, respectively), we investigate the sensitivity of our results to a variety of dust attenuation laws.

Adopting alternative nebular dust prescriptions, such as the \citet{2020ApJ...902..123R} law from the MOSDEF survey or the \citet{2026ApJ...999...15R} law from the AURORA survey, recovers the same mass-dependent SFMS scatter behavior as that derived when using the \citet{1989ApJ...345..245C} law. In addition, the overall SFMS scatter does not vary significantly with the choice of adopted dust law. Given that the \citet{2020ApJ...902..123R} and \citet{2026ApJ...999...15R} curves are derived from deep observations of star-forming galaxies at $z\geq 1.4$, mirroring the sample in this work, their consistency with the \citet{1989ApJ...345..245C} law in deriving the inferred SFMS scatter suggests that the result of a mass-dependent scatter is robust against a reasonable range of nebular dust law prescriptions.

If one corrects the emission-line fluxes using the same dust law that was adopted to correct the stellar light during SED fitting (i.e., a choice between SMC and \citet{2000ApJ...533..682C}), the formally measured mass-dependent scatter is absent in all but the $1.4 < z < 2.7$ bin. However, these particular dust laws are not shown to be appropriate choices to correct nebular emission lines for dust \citep[e.g.,][]{2018ApJ...853...56R,2020ApJ...902..123R}, and are thus of limited use in the context of exploring a reasonable range of plausible choices for the nebular dust correction. As such, our result of a mass-dependent H$\alpha$ SFMS scatter is robust against a range of reasonable choices in nebular dust law.

When adopting the \citet{2026ApJ...999...15R} curve, the H$\alpha$ SFMS normalization surpasses that of the UV SFMS in the $2.7<z<4$ bin due to the higher $R_V$ of this dust law compared to \citet{1989ApJ...345..245C}. As a result, the inferred maximum redshift for which the H$\alpha$ SFMS normalization is lower than that of the UV SFMS decreases when adopting this more recent dust prescription. However, when adopting the \citet{2020ApJ...902..123R} curve, the H$\alpha$ SFMS normalization does not change significantly.

The redshift dependence of $\sigma_{\rm int, H\alpha}$ also shows some sensitivity to the choice of nebular dust attenuation curve. When adopting a \citet{2020ApJ...902..123R} or \citet{2026ApJ...999...15R} curve, the value of $\sigma_{\rm int, H\alpha}$ increases by 0.05 dex in the $6<z<7$ range while only increasing by a maximum of 0.01 dex for the $1.4<z<4$ samples, thereby weakening the trend of decreasing $\sigma_{\rm int,H\alpha}$ with redshift. Interestingly, the redshift-dependent trend remains when analyzing $\sigma_{\rm mass, 2}$ with respect to redshift. Ultimately, the robustness of this result is likely most sensitive to small number statistics, and larger sample numbers at $z\gtrsim 6$ will be needed to confirm the presence of a trend with redshift. In addition, when adopting a \citet{2026ApJ...999...15R} dust curve, the redshift evolution of the sSFR (parametrized as $\rm sSFR \propto(1+z)^\gamma$), for the H$\alpha$ sSFR increases to $\gamma=2.28^{+0.16}_{-0.16}$. When adopting the \citet{2020ApJ...902..123R} curve, the value is $\gamma = 1.91^{+0.16}_{-0.16}$, in close agreement with the \citet{1989ApJ...345..245C}-based value.

We note that when varying the nebular dust attenuation curve between \citet{1989ApJ...345..245C}, \citet{2020ApJ...902..123R}, and \citet{2026ApJ...999...15R}, the key indicators of bursty star formation remain, including a larger intrinsic H$\alpha$ SFMS scatter ($\sigma_{\rm int,H\alpha}$) than the UV intrinsic scatter ($\sigma_{\rm int,UV}$), a mass dependence in the H$\alpha$ SFMS scatter, and a significant fraction of galaxies being characterized by H$\alpha$/UV ratios that are inconsistent with a constant SFH. The resulting SFMS fits adopting these various dust prescriptions are presented in Table \ref{tab:dust_law_exercise} of Appendix \ref{sec:dust_curves_app}.

\section{Conclusions} \label{sec:conclusions}

In this work, we have analyzed the star-formation properties of a sample of 659 galaxies at $1.4<z<9$ with rest-optical spectroscopic and photometric observations from the JADES DR3 and AURORA surveys. We have measured UV and Balmer-line luminosities to calculate SFRs on both long ($\sim$50--100 Myr) and short ($\sim$5--10 Myr) timescales, applying dust corrections to each measurement independently. We also compare our sample to the photometric sample in JADES \citepalias{2024MNRAS.535.2998S} and 3D-HST \citep{2012ApJS..200...13B,2014ApJS..214...24S}, finding that our spectroscopic sample is representative above $10^{8.5}\rm\ M_\odot$, with the exception of the $4<z<5$ bin that lacks bright galaxies at $10^{10}\rm\ M_\odot$. We present the following results:

\begin{enumerate}
    \item When comparing SFR estimates based on observational tracers ($\rm SFR_{UV}$, $\rm SFR_{H\alpha}$) with {\sc prospector}-based SFR estimates using a non-parametric SFH ($\rm SFR_{10}$, and $\rm SFR_{100}$), we find the strongest correlation between $\rm SFR_{UV}$ and $\rm SFR_{100}$, and between $\rm SFR_{H\alpha}$ and $\rm SFR_{10}$ respectively. Though in detail, the timescales probed by H$\alpha$ and UV light are likely shorter than 10 Myr and 100 Myr, respectively, we see that H$\alpha$ and UV light are tracing shorter and longer timescale changes to galaxy SFRs.
    \item In the redshift ranges $1.4<z<6$ (excluding the non-representative $4<z<5$ bin), we find an increasing $\sigma_{\rm int,H\alpha}$ with decreasing stellar mass and tentative evidence of further increased scatter below $10^{8.5}\rm\ M_\odot$, inferred from the notably higher sSFR measurements at these masses. Above $10^{8.5}\rm\ M_\odot$, we do not find mass dependence in $\sigma_{\rm int,UV}$. The mass dependence of $\sigma_{\rm int,H\alpha}$ that we find is consistent with models and simulations that predict highly bursty SFHs in low-mass dark matter haloes due to strong stellar feedback modulating the SFR on short timescales \citep[e.g.,][]{2015MNRAS.451..839D,2017MNRAS.466...88S,2018MNRAS.478.1694M,2020MNRAS.498..430I,2022MNRAS.511.3895F,2023MNRAS.525.2241H}. See Appendix D for a discussion of nebular dust curve uncertainties.
    \item We measure a decreasing $\sigma_{\rm int,H\alpha}$ and $\sigma_{\rm int,UV}$ with increasing redshift, with some dependence of $\sigma_{\rm int,H\alpha}(z)$ on the choice of nebular dust curve. If robust, this trend is consistent with the interpretation that at low redshifts, long-timescale effects on galaxy SFHs are significant, while at high redshift, these effects are less pronounced, leading to a lower overall scatter \citep{2025arXiv250300106M}. However, larger samples at $z\gtrsim 6$ will serve to confirm the robustness of this result.
    \item We analyzed the $\rm L_{H\alpha}/\nu L_{\nu,1600}$ ratios and $\rm SFR_{10}/SFR_{100}$ values inferred from SED fitting, finding that 41--60\% of the galaxy population is poorly described by a constant SFH. We also find tentative evidence for rising $\rm L_{H\alpha}/\nu L_{\nu,1600}$ ratios with increasing redshift. Though the increasingly restrictive limiting line flux sensitivity with redshift likely contributes to this trend, we offer several physical explanations, including increased $\alpha$ enhancement and lower metallicities at high redshifts, more smoothly rising SFHs at early times, or increasingly frequent burst episodes. We investigate how the frequency of bursts evolves with redshift based on the slope of the distribution of galaxies in the $\rm L_{H\alpha}/\nu L_{\nu,1600}$ vs. $\rm\Delta\log(L_{H\alpha})$ plane. We find a slightly steeper slope in this parameter space in our highest redshift bins, consistent with shorter-timescale bursts being more common at high redshift.
    \item We examined the redshift evolution of the sSFR at $10^{9.11}\rm\ M_\odot$ with the functional form $\rm sSFR \propto (1+z)^\gamma$. We found that $\gamma=1.89^{+0.16}_{-0.15}$ for the H$\alpha$ sSFR, and $\gamma=1.36^{+0.13}_{-0.13}$ for the UV sSFR. We also found a higher normalization for the UV-based sSFRs (and therefore in the UV SFMS) than the H$\alpha$-based sSFRs at $z<4$, a prediction consistent with bursty star formation. This result, however, depends somewhat on the nebular dust attenuation law, and may also be probing rising average SFHs.
\end{enumerate}

This work highlights the rich insights into the growth of galaxy formation and evolution in the early universe provided by the observations from the JWST. To complement the deep NIRCam observations being taken of galaxies at high redshifts, a larger, representative sample of galaxies will need to be observed with deep spectroscopy, probing very low SFRs and stellar masses in order to better characterize the low-sSFR phases of galaxy growth in the early Universe.


\section*{Acknowledgments}
We would like to acknowledge the JADES team for their efforts in designing, executing, and making public their JWST/NIRSpec and JWST/NIRCam survey data. We would also like to thank Charlotte Simmonds for kindly sharing her photometric catalog, allowing us to evaluate the robustness of the sample presented in this work. We also acknowledge support from NASA grants JWST-GO-01914 and JWST-GO-03833, and NSF AAG grants 2009313, 2009085, 2307622, and 2307623. This work is based on observations made with the NASA/ESA/CSA James Webb Space Telescope as well as the NASA/ESA Hubble Space Telescope. The data were obtained from the Mikulski Archive for Space Telescopes at the Space Telescope Science Institute, which is operated by the Association of Universities for Research in Astronomy, Inc.,
under NASA contract NAS5-03127 for JWST and NAS 5–26555 for HST. Data were also obtained from the DAWN JWST Archive maintained by the Cosmic Dawn Center. The specific observations analyzed can be
accessed via \dataset[doi:10.17909/8tdj-8n28]{https://archive.stsci.edu/doi/resolve/resolve.html?doi=10.17909/8tdj-8n28},
\dataset[doi:10.17909/gdyc-7g80]{https://archive.stsci.edu/doi/resolve/resolve.html?doi=10.17909/gdyc-7g80}, \dataset[doi:10.17909/fsc4-dt61]{https://archive.stsci.edu/doi/resolve/resolve.html?doi=10.17909/fsc4-dt61}, and \dataset[doi:10.17909/evwk-m381]{https://archive.stsci.edu/doi/resolve/resolve.html?doi=10.17909/evwk-m381}. This work used computational and storage services associated with the Hoffman2 Cluster which is operated by the UCLA Office of Advanced Research Computing’s Research Technology Group.

%

\vspace{35mm}
\facilities{\it JWST, HST}


\software{ Astropy \citep{2013A&A...558A..33A,2018AJ....156..123A,2022ApJ...935..167A}, \texttt{emcee} \citep{2013PASP..125..306F}, \texttt{jwst} \citep{2023zndo...7577320B}, \texttt{msaexp} \citep{2022zndo...7299500B}, \texttt{scipy} \citep{2020NatMe..17..261V}, {\sc prospector} \citep{2021ApJS..254...22J},
\texttt{scikit-learn} \citep{scikit-learn},
\texttt{lmfit} \citep{https://doi.org/10.5281/zenodo.15014437},
\texttt{PyNeb} \citep{2015A&A...573A..42L},
\texttt{linmix}\citep{2007ApJ...665.1489K}
}


\appendix

\section{Treatment of extended objects} \label{sec:ext_objs}

Because the NIRSpec/MSA observations were taken with a 3-point dither pattern, galaxies of comparable size or larger than the $\sim$0\farcs5 offsets were affected by self-subtraction when the observations at different dither positions were combined. Thus, a customized data reduction approach was required to deal with these targets.

The first step in this process was to identify galaxies that significantly suffered from self-subtraction. We identified these galaxies using a random forest classifier (specifically, the \texttt{RandomForestClassifier} module of the \texttt{sklearn.ensemble} Python package; \citet{scikit-learn}). We trained the random forest on a subset of our sample consisting of 156 galaxies, classifying each of them by eye into three categories: non self-subtracting, self-subtracting, and subtracted by an adjacent object. Of the 156 galaxies, we identified 97 galaxies as having no self-subtraction in their spectra, 50 galaxies with self-subtraction in their spectra due to an extended size, and 9 galaxies with self-subtraction due to a bright adjacent galaxy. Using this sample as our training set, we trained a random forest model on the following features: redshift, F444W half-light radius, Kron radius, semi-major axis length from the detection image, semi-minor axis length from the detection image, light profile full-width half-maximum, a neighboring star contamination flag, and a neighboring galaxy contamination flag. All of these quantities are available in the JADES photometric catalog with data and descriptions provided on MAST\footnote{https://archive.stsci.edu/hlsp/jades}.
We performed 1,000 train-test splits on the sample, each time taking a random 30\% of the sample to be the test set. We found that our random forest was able to identify extended, self-subtracting galaxies with a recall rate of 84\% and a specificity of 92\%. Overall, we identified 115 extended galaxies that suffered from self-subtraction.

After identifying the self-subtracting galaxies, we re-reduced their NIRSpec data consistently with \citet{2025A&A...693A..60H}, reducing the data using the reference files from the \texttt{jwst\_1298.pmap} CRDS context. We combined only the outer two dither positions so as to mitigate the effects of self subtraction from the small 3-point dither offsets.

\section{Flux calibration summary plots} \label{sec:flux_cal_app}

\begin{figure}
    \centering
    \includegraphics[width=0.5\linewidth]{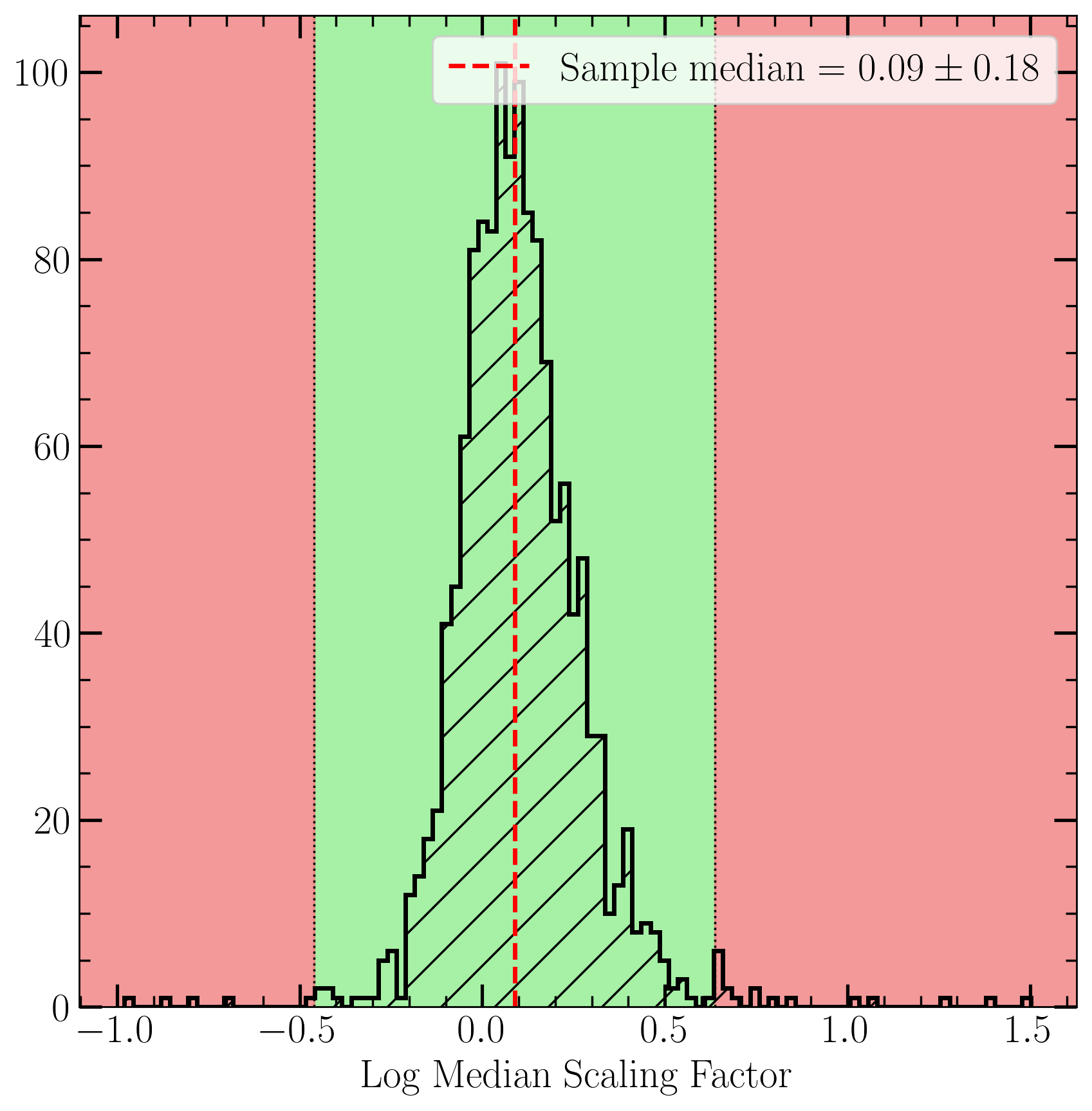}
    \includegraphics[width=0.5\linewidth]{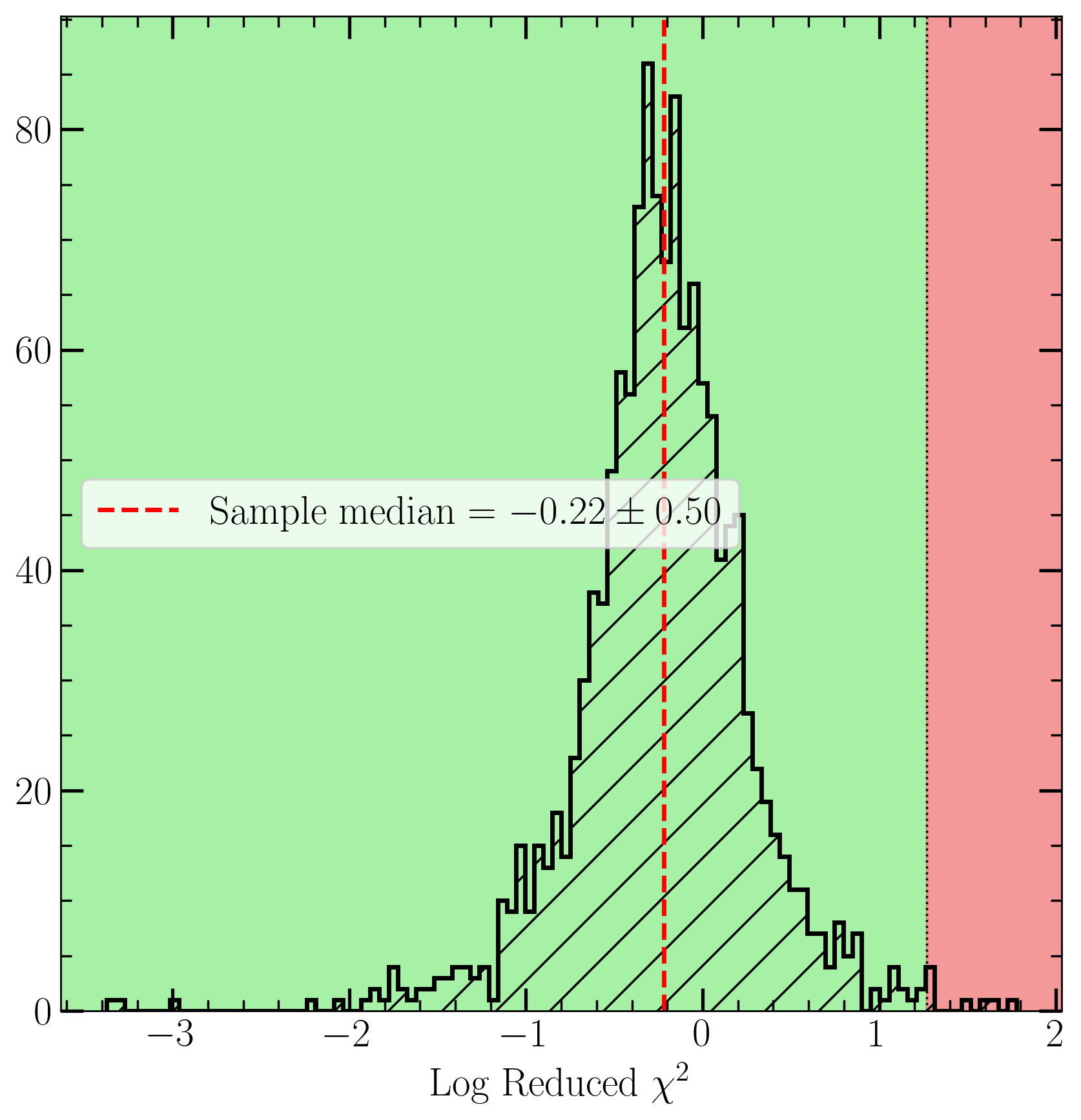}
    \caption{{\it Top panel}: Distribution of the median value of the scaling polynomial (equation \ref{eq:fluxcal}) for each spectrum. The green region illustrates the $\pm3\sigma$ region around the sample median. Objects outside of this range (in the red regions) were removed from the analysis. {\it Bottom panel}: Distribution of reduced $\chi^2$ values of the scaling polynomial (equation \ref{eq:fluxcal}) fit to the photometry. Shaded in red is the range of $\chi^2$ values that lie $>$3$\sigma$ above the sample median. Objects in this range were rejected and removed from the analysis.}
    \label{fig:flux_cal_fig}
\end{figure}

For each object, we plot on a log scale the median value of the scaling polynomial described in Section \ref{sec:fluxcal} and the reduced $\chi^2$ value in Figure \ref{fig:flux_cal_fig}. We only consider objects whose median scaling factors and reduced $\chi^2$ values fall within the green portions of Figure \ref{fig:flux_cal_fig}. The green portions highlight objects whose log median scaling factors are within $\pm3\sigma$ of the sample median, and objects whose reduced $\chi^2$ are less than $+3\sigma$ away from the sample median.

\section{Calculating the limiting sensitivity curves}
\label{sec:sensitivity_curves_app}

We determine the limiting sensitivity by calculating the minimum line flux observed at  $>$3$\sigma$ significance in each redshift bin, excluding the AURORA targets and the deep JADES tiers, since they represent much longer exposure times than the majority of the sample. The estimated limiting line flux across each of the gratings is reported in Section \ref{sec:linefitting}. To convert the limiting line sensitivity to a limiting $\rm L_{H\alpha}$, we ascribe the limiting sensitivity to the H$\beta$ flux and multiply by 2.79 (the theoretical H$\alpha$/H$\beta$ ratio for Case B recombination). We ascribe the sensitivity limit to H$\beta$ rather than H$\alpha$ because we require the H$\alpha$ luminosity to be {\it dust corrected} with the detection of at least two lines, and H$\beta$ is the next brightest recombination line after H$\alpha$ in the absence of dust. Then, in order to convert this limiting line sensitivity into a limiting $\rm L_{H\alpha}/\nu L_{\nu,1600}$ ratio as a function of stellar mass, we adopted our best-fit relationships between $\rm \nu L_{\nu,1600}$ and stellar mass from equation \ref{eq:sfms}, assuming the SMC+0.28Z$_\odot$ conversion from SFR to $\rm \nu L_{\nu,1600}$ and best-fit parameters from Table \ref{tab:sfms_table}. The dashed gray lines in Figure \ref{fig:ha_uv_ratio} indicate how the fitted SFMS intrinsic scatter affect our estimate of the limiting $\rm L_{H\alpha}/\nu L_{\nu,1600}$ boundary. The main purpose of this shaded region is to guide the eye toward features in the sample that may indicate incompleteness driven by exposure time limits.

\section{Analysis of varying the nebular attenuation curve}
\label{sec:dust_curves_app}

In this section of the Appendix, we present our re-calculated values of the top panel of Table \ref{tab:sfms_table}, varying only the adopted nebular dust attenuation law. We have tested the nebular attenuation laws of \citet{2020ApJ...902..123R}, \citet{2026ApJ...999...15R}, as well as adopting a choice between the \citet{2000ApJ...533..682C} and \citet{2003ApJ...594..279G} (SMC) stellar attenuation curves, consistent with the dust law applied to the stellar component of the SED. The choice of dust law applied to the nebular lines does not significantly affect the overall SFMS scatter. The existence of a mass dependence of the SFMS scatter shows some dust-law dependence, with the stellar dust attenuation laws (Calzetti/SMC) suggesting no significant $\sigma_{\rm H\alpha}$ mass dependence, while nebular dust curves derived from constraints based on $z>1$ star-forming galaxies suggest a $\sigma_{\rm H\alpha}$-$M_*$ anticorrelation, similar to that found when adopting \citet{1989ApJ...345..245C} to correct the nebular lines. Given that the stellar dust attenuation curves are not shown to be appropriate choices to apply to the correction of nebular emission lines \citep[e.g.,][]{2018ApJ...853...56R,2020ApJ...902..123R}, we argue that the Calzetti/SMC dust laws do not constitute plausible alternative nebular dust prescriptions, and that the SFMS mass dependence that we measure is robust to reasonable choices of nebular dust attenuation law.

\begin{deluxetable*}{c|ccccccc}
\label{tab:dust_law_exercise}
\caption{Best-fit parameters for the H$\alpha$-based SFMS, adopting different nebular dust attenuation curves.}
\tablehead{
 \colhead{ $ z$ bin} & \colhead{$ \alpha$} & \colhead{$ \beta_N$} & \colhead{$\mathbf \sigma_{\rm int}$} & \colhead{$ \sigma_{\rm mass,1}$$\tablenotemark{\scriptsize a}$} & \colhead{$\sigma_{\rm mass,2}$$\tablenotemark{\scriptsize b}$} & \colhead{$ \sigma_{\rm mass,3}$$\tablenotemark{\scriptsize c}$} & \colhead{$ \rm N_{gal}$} \\ \hline \multicolumn{8}{c}{ AURORA; \citet{2026ApJ...999...15R}}}
\startdata
\hline
$ 1.4 < z < 2.7$ & $ 0.85^{+0.06}_{-0.06}$ & $ 0.30^{+0.03}_{-0.03}$ & $ 0.37^{+0.02}_{-0.02}$ & $ 0.76^{+0.04}_{-0.04}$ & $ 0.40^{+0.02}_{-0.01}$ & $ 0.32^{+0.01}_{-0.02}$ &  165 \\
$ 2.7 < z < 4$ & $ 0.85^{+0.06}_{-0.06}$ & $ 0.62^{+0.03}_{-0.03}$ & $ 0.35^{+0.02}_{-0.02}$ & $ 0.56^{+0.03}_{-0.03}$ & $ 0.38^{+0.02}_{-0.01}$ & $ 0.30^{+0.02}_{-0.02}$ &  154 \\
$ 4 < z < 5^*$ & $ 0.76^{+0.26}_{-0.24}$ & $ 0.92^{+0.07}_{-0.07}$ & $ 0.36^{+0.05}_{-0.04}$ & $ 0.33^{+0.04}_{-0.04}$ & $ 0.37^{+0.02}_{-0.02}$ & $ 0.47^{+0.06}_{-0.06}$ &  66 \\
$ 5 < z < 6$ & $ 0.92^{+0.29}_{-0.28}$ & $ 1.10^{+0.08}_{-0.08}$ & $ 0.38^{+0.07}_{-0.06}$ & $ 0.49^{+0.03}_{-0.03}$ & $ 0.36^{+0.04}_{-0.04}$ & $ 0.06^{+0.11}_{-0.05}$ &  31 \\
$ 6 < z < 7$ & $ 1.12^{+0.27}_{-0.26}$ & $ 1.14^{+0.09}_{-0.09}$ & $ 0.27^{+0.10}_{-0.08}$ & $ 0.73^{+0.09}_{-0.09}$ & $ 0.21^{+0.06}_{-0.07}$ & $ 0.37^{+0.12}_{-0.12}$ &  17 \\
\hline 
\multicolumn{8}{c}{ MOSDEF; \citet{2020ApJ...902..123R}} \\ 
\hline
$ 1.4 < z < 2.7$ & $ 0.85^{+0.05}_{-0.05}$ & $ 0.27^{+0.03}_{-0.03}$ & $ 0.36^{+0.02}_{-0.02}$ & $ 0.76^{+0.04}_{-0.04}$ & $ 0.39^{+0.01}_{-0.01}$ & $ 0.32^{+0.01}_{-0.02}$ &  177 \\
$ 2.7 < z < 4$ & $ 0.85^{+0.06}_{-0.06}$ & $ 0.52^{+0.03}_{-0.03}$ & $ 0.34^{+0.02}_{-0.02}$ & $ 0.58^{+0.02}_{-0.02}$ & $ 0.37^{+0.01}_{-0.01}$ & $ 0.32^{+0.02}_{-0.02}$ &  171 \\
$ 4 < z < 5^*$ & $ 0.77^{+0.21}_{-0.21}$ & $ 0.78^{+0.06}_{-0.06}$ & $ 0.32^{+0.04}_{-0.03}$ & $ 0.36^{+0.03}_{-0.03}$ & $ 0.33^{+0.02}_{-0.02}$ & $ 0.44^{+0.04}_{-0.04}$ &  72 \\
$ 5 < z < 6$ & $ 0.88^{+0.26}_{-0.25}$ & $ 0.95^{+0.07}_{-0.07}$ & $ 0.34^{+0.06}_{-0.05}$ & $ 0.48^{+0.03}_{-0.03}$ & $ 0.32^{+0.03}_{-0.03}$ & $ 0.16^{+0.07}_{-0.06}$ &  31 \\
$ 6 < z < 7$ & $ 0.92^{+0.25}_{-0.24}$ & $ 0.98^{+0.08}_{-0.08}$ & $ 0.27^{+0.09}_{-0.07}$ & $ 0.55^{+0.06}_{-0.06}$ & $ 0.23^{+0.04}_{-0.05}$ & $ 0.27^{+0.11}_{-0.12}$ &  17 \\
\hline 
\multicolumn{8}{c}{ Calz/SMC; \citet{2000ApJ...533..682C,2003ApJ...594..279G}} \\
\hline
$ 1.4 < z < 2.7$ & $ 0.89^{+0.05}_{-0.05}$ & $ 0.19^{+0.03}_{-0.03}$ & $ 0.35^{+0.02}_{-0.02}$ & $ 0.81^{+0.04}_{-0.04}$ & $ 0.38^{+0.01}_{-0.01}$ & $ 0.31^{+0.01}_{-0.02}$ &  170 \\
$ 2.7 < z < 4$ & $ 0.87^{+0.06}_{-0.06}$ & $ 0.40^{+0.03}_{-0.03}$ & $ 0.34^{+0.02}_{-0.02}$ & $ 0.61^{+0.02}_{-0.02}$ & $ 0.35^{+0.01}_{-0.01}$ & $ 0.33^{+0.02}_{-0.02}$ &  172 \\
$ 4 < z < 5^*$ & $ 0.69^{+0.21}_{-0.21}$ & $ 0.64^{+0.06}_{-0.06}$ & $ 0.34^{+0.04}_{-0.03}$ & $ 0.33^{+0.03}_{-0.02}$ & $ 0.33^{+0.01}_{-0.02}$ & $ 0.41^{+0.04}_{-0.04}$ &  71 \\
$ 5 < z < 6$ & $ 0.83^{+0.21}_{-0.21}$ & $ 0.82^{+0.06}_{-0.06}$ & $ 0.29^{+0.05}_{-0.04}$ & $ 0.44^{+0.03}_{-0.02}$ & $ 0.26^{+0.03}_{-0.03}$ & $ 0.25^{+0.08}_{-0.05}$ &  31 \\
$ 6 < z < 7$ & $ 0.84^{+0.20}_{-0.20}$ & $ 0.87^{+0.06}_{-0.07}$ & $ 0.22^{+0.07}_{-0.06}$ & $ 0.45^{+0.05}_{-0.05}$ & $ 0.20^{+0.04}_{-0.04}$ & $ 0.17^{+0.14}_{-0.17}$ &  17 \\
\enddata
\tablenotetext{a}{ Intrinsic scatter in the SFMS for galaxies at $\log(M_*/M_\odot) \leq 8.5$. Note that this mass range is below the mass representativeness limit of our sample}
\tablenotetext{b}{ Intrinsic scatter in the SFMS for galaxies at $8.5 < \log(M_*/M_\odot) < 9.5$.}
\tablenotetext{c}{ Intrinsic scatter in the SFMS for galaxies at $\log(M_*/M_\odot) \geq 9.5$.}
\tablenotetext{*}{ We note that the sample in the $4<z<5$ bin lacks UV-bright objects at $\sim$$10^{10}\rm\ M_\odot$, biasing the SFMS slope to low values. We advise caution when interpreting results in this redshift bin.}\end{deluxetable*}

\bibliography{sample631}{}
\bibliographystyle{aasjournal}



\end{document}